\documentstyle[12pt,epsf,twoside]{article}
\begin{document}
\font\names=cmbx10 scaled\magstep1

\hspace{11cm}{HD-THEP-95-05}
\vspace{3cm}
\begin{center}
{\large\bf Phase Diagram of Superconductors from}\\
\vspace{.3cm}
{\large\bf Non-Perturbative Flow Equations}\\
\vspace{1cm}
{B. Bergerhoff,\footnote{E-mail: B.Bergerhoff@thphys.uni-heidelberg.de}
 F. Freire,\footnote{E-mail: F.Freire@thphys.uni-heidelberg.de}
 D. Litim,\footnote{E-mail: cu9@ix.urz.uni-heidelberg.de; address from
Oct. 1st: Theoretical Physics Group, Imperial College, Prince Consort Road,
London SW7 2BZ, U.K. (E-mail: D.Litim@ic.ac.uk)}}\\
{S. Lola
\footnote{E-mail: S.Lola@thphys.uni-heidelberg.de; address from
Oct. 1st: Department of Theoretical Physics,
University of Ioannina, 451 10 Ioannina, Greece.}
 and C. Wetterich
\footnote{E-mail:
C.Wetterich@thphys.uni-heidelberg.de}}\\
\medskip
{Institut  f\"ur Theoretische Physik, Universit\"at Heidelberg,}\\
{Philosophenweg 16, 69120 Heidelberg, Germany}\\
\medskip

\vspace{.3cm}
(final version to appear in Phys.~Rev.~B)
\vspace{.7cm}

\large\abstract{The universal behaviour of superconductors near the
phase transition is described
by the three-dimensional field theory of scalar quantum electrodynamics. We
approximately solve the model with the help of non-perturbative flow
equations.
A first- or second-order phase transition is found depending on the relative
strength of the scalar versus the gauge coupling. The region of a
second-order phase transition is governed by a fixed point of the flow
equations with associated critical exponents. We also give an approximate
description of the tricritical behaviour and briefly discuss the crossover
relevant for the onset of scaling near the critical temperature. Final
confirmation of a second-order transition for strong type-II superconductors
requires further analysis with extended truncations of the flow equations.}
\end{center}

\thispagestyle{empty}

\setcounter{page}{0}
\vfill\eject


\begin{section}{\names Introduction}

Superconductors of type-I and type-II are distinguished by the
ratio of the photon
mass, $M$, and the scalar mass, $m$, in the superconducting phase at low
temperature. These masses correspond to two relevant length scales,
namely the
London penetration depth, $M^{-1}$, and the coherence length, $m^{-1}$. By
convention a superconductor is called of type-I (II) if
$M^2/m^2 {\:\hbox to -0.2pt{\lower2.5pt\hbox{$\sim$}\hss}
           {\raise3pt\hbox{$>$}}\:} ({\:\hbox to -0.2pt{\lower2.5pt
        \hbox{$\sim$}\hss}
           {\raise3pt\hbox{$<$}}\:})~1$.
Only in type-II superconductors magnetic vortices can occur.
The
question arises whether type-I and type-II superconductors are
distinguished not
only by their low temperature properties but also by a different
behaviour near
the critical temperature, $T_c$, of the normal-to-superconductor
phase transition.

The critical behaviour near $T_c$ is thought to be described by
the field theory
of a charged scalar particle coupled to a photon.
Whereas the photon represents the electromagnetic degrees
of freedom, the scalar is associated with collective
excitations, as the Cooper pairs in standard superconductors.
The most relevant interactions are determined by the gauge
coupling (effective electric charge) $\bar{e}$ and the
quartic scalar self-coupling $\bar{\lambda}$.
In turn, the masses of the gauge boson and the scalar
in the superconducting phase are proportional to
$\bar{e}$ and $\sqrt{\bar\lambda}$, {\it {i.e.}~}
$M^2/m^2=\bar e^2/\bar\lambda$.
Equilibrium properties in classical statistical systems \footnote{We
do not deal here with
possible
quantum statistics effects in the extreme low temperature limit $T\to0$.}
are given by Euclidean field theories. In case of
the superconductor we assume that
three-dimensional
scalar quantum electrodynamics (QED) provides a good
description of the physics whenever
the composite character
of the scalar field is negligible and
when the system is essentially rotation invariant. These conditions are
fulfilled for $T$ sufficiently near $T_c$.
Although the Ginzburg-Landau theory suggests a second-order
transition for all
values of $\bar e^2$ and $\bar\lambda$, it has been proposed
for a long time \cite{HLM}
that good type-I superconductors ($\bar\lambda \ll \bar e^2$)
exhibit a first-order
transition. The discontinuity in the order parameter \footnote{Strictly
speaking the local gauge symmetry is never spontaneously broken. With the
gauge fixing we are using this picture is nevertheless very useful.} -- the
vacuum expectation
value of the charged scalar field -- is induced by fluctuations of the gauge
field. On the other hand, there have been arguments
based on analogies with various related models that the phase transition of
a good type-II superconductor~ ($\bar\lambda \gg \bar e^2$)
may be second-order \cite{{DH},{noepsilon},{rad}}.
In \cite{HLM} it was originally conjectured
that the transition in type-II superconductors
is also a fluctuation induced first-order one. Such a result is thought to be
an artefact of the $\epsilon$-expansion when stretched to three dimensions as
remarked by several authors \cite{{noepsilon},{rad}}.
Direct computations within the field theory of a charged scalar field coupled
to a photon have not provided conclusive results in the past.
Strong infrared divergences appearing in the loop expansion at the critical
temperature of a second-order transition are the reason keeping this long
standing problem unsolved.

The key problem for the understanding of a weak first- or a
second-order transition is the presence of long range fluctuations.
For given ``microscopic interactions'' at some length scale
$\Lambda^{-1}$ the long range fluctuations lead to different
effective interactions at some larger length scale
$k^{-1}>\Lambda^{-1}$. In particular, the effective couplings
$\bar e^2$ and $\bar\lambda$ will depend on $k$. For an understanding of
a weak first-order transition with a small photon mass $M_c$ at the
critical temperature $T_c$ one needs the couplings at $k\simeq M_c$.
The critical behaviour at a second-order transition even
involves the limit $k\rightarrow 0$. (For first-order transitions
one may also take $k\rightarrow 0$, since effective couplings
remain constant once $k$ is smaller than the lightest mass.)
The scale variation (or ``running'') of the couplings $\bar{e}(k)$ and
$\bar{\lambda}(k)$ is described by flow equations or renormalization
group equations.

We parametrize the theory by the short distance couplings
$\bar{e}^2(\Lambda)$ and $\bar{\lambda}(\Lambda)$.
If there is a range in
$\bar e^2$ and $\bar\lambda$ for which the phase transition
is second-order, the
renormalization group flow of the corresponding renormalized dimensionless
couplings ${e^2}(k)$ and $\lambda(k)$ should be attracted by
a fixed point. A firm
establishment of a second-order transition for large $\bar\lambda$
therefore  requires
finding this ``second-order fixed point'' which must be infrared
stable for
both $\lambda$ and ${e^2}$. Furthermore, for a continuous transition
the short distance values of $\bar\lambda$ and $\bar e^2$ must lie
within the domain of attraction of this fixed point.
On the other hand, if the
phase transition is first-order for small $\bar\lambda$
there must be a line of separation
in the $\bar\lambda$-$\bar e^2$ plane between the first- and
second-order behaviour. At the
critical temperature the trajectories of $\lambda$ and ${e^2}$
which start on this
separation line flow towards a tricritical fixed point. The tricritical fixed
point has one direction that is infrared stable and another one unstable: the
instability corresponds to the flow towards either the first- or the
second-order region. We should emphasize that
for $\bar\lambda$ and $\bar e^2$ sufficiently
close to the separation line there is not much distinction, in practice,
between
the first- and second-order transition. The behaviour on both sides of the
separation line is governed by the tricritical fixed point -- or the
crossover from the gaussian fixed point ($\bar\lambda=\bar e^2=0$) to the
tricritical point -- for temperatures near
to, but not extremely close to, the critical temperature. The first-order
transition is very weak in this case and the distinction from a second-order
transition becomes possible only on scales which are tiny compared to $T_c$.
Consequently numerical simulations \cite{DH} often
cannot tell a second- from a weak
first-order transition. It would require an observation of the crossover from
the tricritical fixed point to the ``second-order fixed point'' in order to
distinguish them.

Using the ``Ginzburg criterion'' for the size of the temperature range around
$T_c$ where fluctuations become important one concludes that experimental
verification of the order of the phase transition seems extremely difficult
for standard low temperature superconductors.
This situation may improve for high
temperature superconductors~\cite{hightc}. Furthermore, it
has been proposed~\cite{gennes}
that the nematic-to-smectic-A transition in liquid crystals
belongs to the same universality class as the superconductor transition.
Here a
second-order phase transition as well as a tricritical behaviour have
been found and critical exponents are
measured~\cite{{nemdata},{newdata}}. A theoretical establishment
of a second-order phase
transition for large $\bar\lambda$ would therefore be useful
also for practical purposes.

In this paper we investigate the phase diagram for superconductors
with the use of a non-perturbative
flow equation~\cite{CW}.  This evolution equation describes the dependence
of the average action $\Gamma_k$~\cite{CW1}
on the ``average scale'' $k$. The
average action corresponds to a coarse grained free energy functional with
$k^{-1}$ the length scale of the coarse graining.
One obtains it by integrating
out all fluctuations with momenta $q^2 > k^2$. For $k\to0$
one recovers the free
energy functional, {\it {i.e.}~} the generating functional
for the 1PI Green functions.
In particular, we will study the scale dependence of the average potential
$U_k(\rho)$ which corresponds to the coarse grained free energy for constant
values of the complex scalar field $\varphi$,
$\rho=|\varphi|^2$. If for $k\to0$ the
minimum of the potential occurs for $\rho_0\neq0$ the gauge symmetry is
spontaneously broken and the model is in the superconducting phase.
When the minimum of $U_k(\rho)$ is at the origin, $\rho_0=0$,
the symmetric phase
-- normal conductor -- is realized. Our flow equation
for $U_k$~ derives from the
exact non-perturbative flow equation \cite{NPB427}
for $\mbox{$\Gamma$}_k$ in an appropriate
truncation.

The exact flow equation for $\mbox{$\Gamma$}_k$
is equivalent to earlier versions
of exact renormalization group equations \cite{polchinski}.
However, it differs
from the ``cut-off action'' used in these previous versions in that
it describes a coarse grained free energy in continuous space.
This allows for an easy and direct treatment of infrared problems.
The flow equation can be derived
\cite{CW}
from a simple identity for the functional integral defining the
field theory once an infrared cut-off term
$\sim \psi R_k \psi$ has been added to the action.
The cut-off should be quadratic in the general quantum fields $\psi$ and
may depend on momentum.
For small momenta $q^2<k^2$ it acts as a mass term $R_k \sim k^2$
and therefore plays the role of an infrared regulator.
The flow equation has the simple
interpretation as a renormalization group improved
one-loop equation
\begin{eqnarray}
\frac{\partial}{\partial k} \Gamma_k[\psi] =
\frac{1}{2} \int \frac{d^dq}{(2\pi)^d} \mbox{tr}\left\{
\left( \Gamma^{(2)}_k[\psi]+R_k\right)^{-1}\frac{\partial}
{\partial k}R_k \right\} \,\, .
\label{new_1}
\end{eqnarray}
Here $\Gamma_k^{(2)}$ is the exact inverse propagator
obtained as the second functional derivative of
$\Gamma_k$ with respect to the fields $\psi$.
For suitable $R_k(q^2)$
not only infrared divergences but also
ultraviolet ones are absent in our approach -- for each
infinitesimal change of
length scale $k^{-1}$ only modes with momentum close to $k$
contribute to the flow.
The couplings parametrizing the interactions contained in $\Gamma_k$
depend on $k$ and include $\bar{e}(k)$ and $\bar{\lambda}(k)$.
The flow equation describes the variation of these couplings
with $k$. The most general form of
$\Gamma_k$ involves infinitely many couplings and an exact
solution of the functional differential equation (\ref{new_1})
is out of reach. In one way or another, one has to reduce
the system to a manageable size and restrict the most
general form of $\Gamma_k$. If such a truncation includes
all interactions relevant for the physical system, one expects
it to provide an approximate solution of (\ref{new_1}).
Unfortunately, such truncations are often not very systematic
in situations where no small parameters are available, as in the present
case for the critical superconductor. Insight into the
dominant physics of the system is probably the best guide for
deriving appropriate truncations under such circumstances,
with the possibility to check the self-consistency of the
results a posteriori by computing if the neglected
interactions remain small in the course of the evolution.
In the present paper we use simple truncations and discuss
their shortcomings at the end. Very encouraging results
for critical scalar field theories, including the
equation of state, have been obtained with similar
truncations
\cite{{critexp},{scalar_AEA}}.
Within our approximations we find a
second-order fixed point as well as a tricritical fixed point. This
allows to draw a phase diagram for superconductors which is, however, not yet
quantitatively reliable in all regions in parameter space.
The present work is closely related to a similar study~\cite{largeN} with an
arbitrary number $N$ of complex scalar fields. In this
case the superconductor~ phase
transition, {\it {i.e.}~} $N=1$, was approached from large $N$.
A brief outlook on the results of this work will be addressed in section 6.

The paper has the following format.
In section 2 we first use the flow equations in their simplest form to
establish the qualitative features of the phase diagram. These flow equations
as well as more elaborate versions of them are derived
as approximations of the exact flow
equation in sections 3 and 4. In section 5
we discuss the parameter range where
the transition is first-order and we turn to the second-order transition in
section 6. Shortcomings of the present approximations and extensions of our
work which are necessary for a firm
establishment of our picture are indicated
in section 7. Finally section 8 contains our conclusions and a discussion
of the results.

\end{section}


\begin{section}{\names Phase Diagram}

The simplest version of the non-perturbative
flow equations will be described in this section.
We will derive them in the next two sections and more refined
approximations will be discussed in
the following ones. For our purposes it is
most convenient to use the dimensionless renormalized couplings

\begin{eqnarray}
{e^2}(k)&=&\frac{\bar e^2}{Z_{\scriptscriptstyle F}(k)k}
= \frac{e^2_{\scriptscriptstyle R}(k)}{k}
\label{dlesscoup1}
\end{eqnarray}

\noindent and

\begin{eqnarray}
\lambda(k)&=&\frac{\bar\lambda(k)}{Z^2_{\varphi}(k)k}
= \frac{\lambda_{\scriptscriptstyle R}(k)}{k}
\label{dlesscoup2}
\end{eqnarray}

\noindent where $Z_{\scriptscriptstyle F}(k)$,
$Z_{\varphi}(k)$ are appropriate wave function
renormalization factors for the gauge and scalar field, respectively.
For standard superconductors $\bar e^2$ is
related to the fine structure constant,
$\alpha\sim1/137$, by $\bar e^2=16\pi\alpha T$
and $e(k)$ is a running effective
dimensionless gauge coupling in three dimensions. Similarly, the quartic
scalar interaction is described by a
dimensionless running coupling $\lambda(k)$.
If the minimum of the average potential occurs
for a non-vanishing value of the
order parameter, $\rho_0(k)$, we relate its location to
another dimensionless parameter

\begin{eqnarray}
\kappa(k)&=&\frac{Z_{\varphi}(k)\rho_0(k)}{k}.
\label{dlesscoup3}
\end{eqnarray}

\noindent The average potential -- related to a coarse
grained free energy for constant
field values -- is hereby approximated near its minimum by

\begin{eqnarray}
U_k(\rho)&=&{1\over2}\bar\lambda(k)(\rho-\rho_0(k))^2
\label{dfullpotexp}\\
\rho&=&\varphi^{\dagger}\varphi \label{rho}.
\end{eqnarray}

\noindent We use a normalization in units of $T$ which
is adapted to a three-dimensional field theory
where $\bar e^2$, $\bar\lambda$ and $\rho_0$
have dimension of mass. For non-vanishing $\kappa$
we may define a running photon
mass $M(k)$ by

\begin{eqnarray}
M^2(k)=2{e^2}(k)\kappa(k)k^2=\frac{2\bar e^2
\rho_0(k)Z_{\varphi}(k)}{Z_{\scriptscriptstyle F}(k)}
\label{photonmass}
\end{eqnarray}

\noindent and similarly for the scalar mass $m(k)$

\begin{eqnarray}
m^2(k)=2\lambda(k)\kappa(k)k^2=
\frac{2\bar\lambda(k)\rho_0(k)}{Z_{\varphi}(k)}.
\label{scalarmass}
\end{eqnarray}
In the phase with spontaneous symmetry breaking the physical masses are
obtained for $k\to0$, {\it {i.e.}~}  $M^2=M^2(0)$, $m^2=m^2(0)$,
$M^2/m^2={e^2}(0)/\lambda(0)$. Moreover, $\rho_0(k)$ reaches
a constant non-vanishing value for $k\to0$.

We want to study the scale dependence
of the average potential between some
initial (high momentum)
scale $\Lambda$ and $k=0$.
The evolution equations for the dimensionless
parameters ${e^2}$, $\lambda$ and
$\kappa$ in their simplest version, with $t=\ln{k/\Lambda}$,
are given by \cite{NPB427}

\begin{eqnarray}
\frac{d \kappa}{d t} &=& -\kappa + \frac{1}{\pi^2}
      ~\frac{e^2}{\lambda} ~\ell_{1}(2 {e^2} \kappa)
        + \frac{1}{4\pi^2} ~\ell_{1}
        +\frac{3}{4\pi^2}~\ell_{1}(2 \lambda \kappa)
\nonumber\\
 &&+ \frac{4}{3\pi^2}~{e^2}\kappa ~
\ell_{1,1}(2 {e^2} \kappa,2 \lambda \kappa)
 - \frac{2}{3\pi^2}~\lambda^2\kappa^2 ~m_{2,2}(2 \lambda \kappa,0)
\label{kappaflow}\\
\nonumber\\
\frac{d \lambda}{d t} &=& -\lambda +
\frac{2}{\pi^2}~e^4 ~\ell_{2}(2{e^2}\kappa)
+ \frac{1}{4\pi^2}~\lambda^2 \ell_{2}
+\frac{9}{4\pi^2}~\lambda^2~\ell_{2}(2 \lambda \kappa)\nonumber\\
&&- \frac{8}{3\pi^2}~{e^2}\lambda~\ell_{1,1}(2 {e^2}\kappa,2 \lambda\kappa)
+ \frac{4}{3\pi^2}~\lambda^3\kappa~m_{2,2}(2 \lambda \kappa,0)
\label{lambdaflow}\\
\nonumber\\
\frac{d {e^2}}{d t} &=& -{e^2}+ \frac{e^4}{6\pi^2}~\ell_ {\rm eff}.
\label{e2flow}
\end{eqnarray}

\noindent The flow equations (\ref{kappaflow}) and (\ref{lambdaflow})
emerge if one evaluates equation (\ref{new_1}) for constant
scalar fields $\varphi$, inserting on the right-hand side
the ansatz (\ref{dfullpotexp}), together with kinetic terms
multiplied by appropriate wave function renormalization
factors $Z_\varphi$ and $Z_{\scriptscriptstyle F}$.
We have already incorporated effects
of the anomalous dimensions corresponding to the $k$-dependence
of $Z_\varphi$ and $Z_{\scriptscriptstyle F}$
(the latter yields (\ref{e2flow}))
which can be computed similarly by evaluating (\ref{new_1})
for appropriate space dependent configurations.
The structure of the flow equations closely resembles perturbative
expressions as obtained, for example, in the $\epsilon$-expansion
\cite{epsilonexp}.
For example, the contribution $\sim e^4$ in the $\beta$-function
for $\lambda$, eq.~(\ref{lambdaflow}), corresponds to the graph displayed
in Fig.~1.
As compared to the $\epsilon$-expansion, the numerical difference
of the constants multiplying the various contributions to the
$\beta$-functions is explained by the fact that we can
compute directly for $d=3$ instead of working in $d=4-\epsilon$
and extrapolating to $\epsilon=1$. In the present truncation
the dominant non-perturbative feature concerns the appearance
of threshold functions together with a flow equation
for $\kappa$, describing the changing of the potential minimum
as $k$ is lowered. This goes beyond an expansion of the
$\beta$-functions in powers of the couplings and is not visible in the
$\epsilon$-expansion or other purely perturbative methods.
The threshold functions $\ell_n(\omega)$, $\ell_{1,1}(\omega_1,\omega_2)$ and
$m_{2,2}(\omega,0)$ depend on the dimensionless masses, $M^2(k)/k^2$ or/and
$m^2(k)/k^2$, and vanish for large values of the argument. They describe the
effective  decoupling of modes with mass larger than $k$. More details on
the threshold functions will be given later. For the moment we only quote the
numerical values of the constants $\ell_{1}=\ell_{1}(0) = 0.886$,
$\ell_{2}=\ell_{2}(0) = \ell_{1,1}(0,0) = 1.04$
and $m_{2,2}(0,0) = 0.412$.
The constant $\ell_ {\rm eff}$ in eq.~(\ref{e2flow}) is thought
to approximate a relatively complicated threshold function in the appropriate
mass range. To simplify we take here the value for zero mass
$\ell_ {\rm eff} = 0.844$.

Our task is now to solve these flow equations
starting at some initial scale $\Lambda$ which we may identify for the present
purpose with $T$ or some scale characteristic for the formation of the scalar
bound state. More precisely, $\Lambda$ is the
scale below which the effective field
theory can be used.
With $Z_{\varphi}(\Lambda)=Z_{\scriptscriptstyle F}(\Lambda)=1$ we find
${e^2}(\Lambda)=16 \pi \alpha T/\Lambda$
such that $e^2(T)$ is independent of $T$.
The temperature  dependence
arises essentially through the temperature dependence of $\rho_0(\Lambda)$.
There is a critical value $\rho_{0,c}(\lambda(\Lambda),{e^2}(\Lambda))$
which corresponds to
the critical temperature. In the vicinity of $T_c$ we can expand

\begin{eqnarray}
\rho_0(\Lambda)=\rho_{0,c}+a(T_c-T) \label{critrho}
\end{eqnarray}

\noindent with $a>0$ a model dependent dimensionless constant.
Solving for $k\to0$ we will find in the superconducting phase

\begin{eqnarray}
\begin{array}{ccccccc}
Z_{\varphi}(k) &  \to &  Z_{\varphi,0} & ,  &
 Z_{\scriptscriptstyle F}(k) & \to &  Z_{{\scriptscriptstyle F},0} \\[1ex]
\rho_0(k) &  \to &  \rho_0 & ,  &    \bar\lambda(k) &  \to &
\bar\lambda_0 \\[1ex]
M^2(k)  & \to &  M^2 & , &     m^2(k) &  \to &  m^2
\end{array}
\label{ktozero}
\end{eqnarray}

\noindent such that ${e^2}$, $\lambda$ and $\kappa$ diverge $\sim k^{-1}$.
On the other side,
in the symmetric phase $\kappa$ will vanish at some finite scale $k_s >0$.
(We will not be concerned much with this phase in the present paper.)
For temperatures
sufficiently below $T_c$ there will only be little running of the parameters
$Z_{\varphi}$, $Z_{\scriptscriptstyle F}$,
$\rho_0$ and $\bar\lambda$ since no long range fluctuations are present.
This corresponds to the range of validity of the
Ginzburg-Landau theory where
the free energy obeys $U_{\Lambda}(\rho) \approx U_0(\rho)$.
In this temperature range
one may therefore associate

\begin{eqnarray}
M^2 &\approx& M^2(\Lambda)
= 2{e^2}(\Lambda)\rho_0(\Lambda)\Lambda \nonumber\\[1ex]
m^2 &\approx& m^2(\Lambda)
= 2\lambda(\Lambda)\rho_0(\Lambda)\Lambda \label{lowtemp}
\end{eqnarray}

\noindent and use this for a phenomenological
determination of $\rho_0(\Lambda)$ and
$\lambda(\Lambda)$ as functions of $T$ for a given superconducting material.

For $T$ in the vicinity of $T_c$ some of the masses will turn out much smaller
than the temperature. The flow of the couplings becomes relevant and can lead
to important modifications of the Ginzburg-Landau theory. In our approximation
the flow of the gauge coupling is particularly simple and eq.~(\ref{e2flow})
is easily solved. One finds an infrared stable fixed point at

\begin{eqnarray}
e^2_\star=\frac{6\pi^2}{\ell_ {\rm eff}}.
\label{fpeff}
\end{eqnarray}
In the vicinity of this fixed point,
$e^2_{\scriptscriptstyle R}(k) = e^2(k)k$
decreases proportional to the scale $k$.
(Of course, once all particle masses are much larger than $k$,
one should put
$\ell_ {\rm eff}=0$ in a more realistic truncation,
the running of $Z_{\scriptscriptstyle F}(k)$
stops and $e^2_{\scriptscriptstyle R}$ becomes constant.)
For ${e^2}=e^2_\star$ the remaining system of evolution equations
(\ref{kappaflow}-\ref{lambdaflow}) can be
solved numerically. The resulting
phase diagram is shown in Fig.~2, with arrows indicating the flow towards the
infrared $(k\to 0)$. The symmetric phase (small $\kappa$)
is separated from the
superconducting phase (large $\kappa$) by a phase transition line.
On this line we
observe the second-order fixed point at $\lambda_\star\approx 20$
and the tricritical
fixed point at $\lambda_\star\approx 0.03$.
The separation between the first- and second-order
behaviour of the transition is indicated by the dashed line which crosses the
tricritical fixed point. The first-order behaviour corresponds here to the
region of small $\lambda$. In the
three-dimensional parameter space $({e^2},\lambda,\kappa)$
the critical line of Fig.~2 extends to a critical surface. The flow of the
trajectories on this surface is shown in Fig.~3.
Here $\kappa=\kappa_c(\lambda,{e^2})$ corresponds
to the critical value for the phase transition. In Fig.~3 we observe the
gaussian fixed point, $e^2_\star=\lambda_\star=0$,
and the Wilson-Fisher fixed point,
$ e^2_\star=0, \lambda_\star\neq0$,
which are both infrared unstable. On the right part of
the phase diagram the flows are directed
towards the second-order fixed point.
On the left part the trajectories reach $\lambda=0$ and
this corresponds to the
region of a first-order transition.
We again indicate the separation line between
the first- and second-order behaviour and the tricritical point on this
line. In summary, the system of flow equations (\ref{kappaflow}-\ref{e2flow})
clearly corresponds to regions of first- and
second-order transitions separated
by tricritical behaviour. A more detailed picture of the separation line for
small values of ${e^2}$ is given in Fig.~4.

The properties of the phase diagram can
easily be understood from the behaviour of
the $\beta$-functions. In Fig.~5 we give
the function $\beta_{\lambda}=\frac{d\lambda}{dt}$ for ${e^2}= e^2_\star$ and
$\kappa=\kappa_c(\lambda,e^2_\star)$.
The two non-trivial zeros correspond to
the unstable tricritical fixed point and the
stable second-order fixed point \footnote{The zero at $\lambda=0$
corresponds to
$\kappa_c\to\infty$ and is an artefact of our truncation.}.
We emphasize that the threshold functions in
eq.~(\ref{kappaflow}) are crucial for the existence of the two fixed points.
Setting the arguments of the threshold functions to zero
the fixed points disappear.
The omission of threshold effects is the main reason why previous perturbative
estimates, {\it {e.g.}~} the
$\epsilon$-expansion, fail to produce a similar phase diagram.

Only for sufficiently
small values of ${e^2}\kappa$ and $\lambda\kappa$
we can neglect the threshold effects
and approximate the threshold functions
by their values at vanishing argument. We
will often call this the linear
approximation -- despite the fact that the flow
equations remain coupled non-linear differential equations.
In particular, one
finds for the running of the ratio $\lambda/{e^2}$
in the vicinity of the gaussian fixed point

\begin{eqnarray}
\beta_{\lambda\over{e^2}}&=&\frac{d}{d t}\left(\frac{\lambda}{{e^2}}\right)=
\frac{1}{{e^2}}\frac{d\lambda}{d t}
-\frac{\lambda}{e^4}\frac{d{e^2}}{d t}\nonumber\\
&=&\frac{{e^2}}{6\pi^2}\left[12~\ell_{2}-(16~\ell_{2}+\ell_ {\rm eff})
\frac{\lambda}{{e^2}}+15~\ell_{2}\left(\frac{\lambda}{{e^2}}\right)^2\right].
\label{betaratio}
\end{eqnarray}
We show $e^{-2} \beta_{\lambda\over{e^2}}$ in Fig.~6
and observe that there is  no zero of
this function. We conclude that $\lambda/{e^2}$ always decreases and that all
trajectories remaining within the validity of the linear approximation lead
to $\lambda=0$ for $k>0$. This region
corresponds therefore to a first-order phase
transition. (The value of $\kappa_c$ is
irrelevant in this approximation. To a
good approximation it is given by $\kappa_c \approx \frac{\ell_{1}}{\pi^2}
(\frac{{e^2}}{\lambda}+1)$ if $\lambda/{e^2}$ is not too small.)
In order to get a quantitative
idea how fast $\lambda/{e^2}$ decreases
we approximate in eq.~(\ref{betaratio}) the term
in brackets by a constant $C$ -- for $C$ taking the minimum value
$C_{\rm min} = 7.58 $
the true running is then
always faster. With

\begin{eqnarray}
\frac{d}{dt}\left(\frac{\lambda}{{e^2}}\right)=\frac{C}{6\pi^2}{e^2}
\label{appbetaratio}
\end{eqnarray}
and eq.~(\ref{e2flow}) we obtain the explicit solution

\begin{eqnarray}
\frac{\lambda(k)}{{e^2}(k)}&=&\frac{\lambda(\Lambda)}{{e^2}(\Lambda)}
+\frac{C}{\ell_ {\rm eff}}
\ln\frac{1-\frac{\ell_ {\rm eff}}{6\pi^2}{e^2}(k)}{1-\frac{\ell_ {\rm eff}}
{6\pi^2}{e^2}(\Lambda)}
\nonumber\\
&\approx&\frac{\lambda(\Lambda)}{{e^2}(\Lambda)}
-\frac{C}{6\pi^2}({e^2}(k)-{e^2}(\Lambda))\nonumber\\
&\approx&\frac{\lambda(\Lambda)}{{e^2}(\Lambda)}
-\frac{C}{6\pi^2}{e^2}(\Lambda)\left(\frac{\Lambda}{k}-1\right)
\label{ratiosolution}
\end{eqnarray}
where the last two lines use ${e^2}\ll 6 \pi^2 / \ell_ {\rm eff}$.
We note that for $\lambda(\Lambda) / e^2(\Lambda)$
substantially smaller than a characteristic value

\begin{eqnarray}
\left(\frac{\lambda(\Lambda)}{{e^2}(\Lambda)}\right)_ {\rm ch} =
\frac{C}{\ell_ {\rm eff}}\approx 15
\label{approxratio}
\end{eqnarray}
the coupling $\lambda(k)$ reaches zero for a value of $k$
where ${e^2}(k)$ remains
much smaller than  $ e^2_\star=6\pi^2/\ell_ {\rm eff}$.
In this region
of parameter space the running of $e^2_{\scriptscriptstyle R}(k)$
does not have a very
strong effect. In a first approximation
$e^2_{\scriptscriptstyle R}(k)$ may be taken as a constant and we expect that
the phase transition is first-order.
A typical scale for the discontinuity
is set by $k_{\rm dis}$ where $\lambda$ vanishes, with

\begin{eqnarray}
\frac{k_ {\rm dis}}{\Lambda} =
\frac{C}{6\pi^2}\frac{e^4(\Lambda)}{\lambda(\Lambda)}.
\label{discontinuity}
\end{eqnarray}
In particular, we may use $\frac{C}{6\pi^2} =
\frac{2 \ell_{2}}{\pi^2} = 0.21 $
for small values of $\lambda(\Lambda)/e^2(\Lambda)$ as a very good
approximation. It is obvious that $k_ {\rm dis}$
decreases strongly with increasing $\lambda(\Lambda)/e^2(\Lambda)$.
For good type-I superconductors
($\lambda(\Lambda)/e^2(\Lambda) \approx 1/10$)
the ratio $k_{\rm dis}/\Lambda =  {\cal O}(1)$ is not a small quantity.
We will discuss this case in more detail in section~5.
Consider next the borderline between type-I and type-II
superconductors
for $\lambda(\Lambda)/e^2(\Lambda) \approx 1$.
Here the discontinuity already becomes small and
difficult to observe.
The behaviour in the vicinity of the critical temperature
is mainly determined by the gaussian fixed point.
The corresponding (approximate) critical exponents
are given by mean field theory.
Going even further into the type-II range one sees that
for $k_ {\rm dis}/\Lambda \approx e^2(\Lambda)/e^2_\star \approx 10^{-2}$
the running of the gauge coupling begins to become an important effect.
This situation corresponds to the characteristic ratio (\ref{approxratio}).
In particular, the quartic coupling $\lambda(k)$
does not reach zero within the validity of the
linear approximation. In this region in parameter space
we locate a possible change to a second-order behaviour.
For good type-II superconductors with
$\lambda(\Lambda)/e^2(\Lambda) \geq 10$ the quartic scalar
coupling $\lambda(k)$
reaches the vicinity of the Wilson-Fisher
fixed point before $e^2_{\scriptscriptstyle R}(k)$
has decreased very significantly.
As a result, the critical behaviour
of these type of superconductors
is dominated by the Wilson-Fisher
fixed point, with exponents of the two-component
Heisenberg model. The crossover to the second-order fixed point
with its associated exponents can be felt only
very close to $T_c$. Due to the
smallness of $e^2(\Lambda)$ for superconductors the
first observation of scaling behaviour for
$T$ near $T_c$ will always detect the
critical exponents of the gaussian or the
Wilson-Fisher fixed point, with a typical
crossover behaviour between the two extreme cases
as $\lambda(\Lambda)/e^2(\Lambda)$ varies.
This situation is independent of the exact nature
of the phase transition,
provided $k_ {\rm dis}/\Lambda$ is small for a first-order transition.
The distinction between a first-
and a second-order transition as well as the
detection of the ``true'' critical behaviour with exponents
different from the ones of the Heisenberg model,
is only possible in a very tiny temperature interval around
$T_c$.
We emphasize that the situation may be quite
different for lattice simulations or experiments
with materials in the same universality class as
normal superconductors, as for example for
the nematic-to-smectic-A transition in liquid crystals.
Here the effective coupling $e^2(\Lambda)$
may take large values, and the critical exponents
associated with the second-order fixed point
could be more easily observed.
Nevertheless, caution is necessary for the interpretation
of observed critical exponents since crossover
behaviour between different fixed points is possible.
This holds in particular for the crossover between
the Wilson-Fisher fixed point and the ``true''
second-order fixed point (cf.~Fig.~3)
since the critical exponents are not largely different
(see section 6) for both cases.

Comparing the above qualitative discussion with Fig.~4
we notice that the approximation (\ref{kappaflow}-\ref{e2flow})
implies a transition to a second-order behaviour for
$\lambda(\Lambda)/e^2(\Lambda) \approx 0.1$ and therefore
much smaller than the characteristic value (\ref{approxratio}).
This is related to the fact that $2e^2\kappa$
reaches one along the corresponding
trajectories before $\lambda(k)$ reaches zero.
The linear approximation therefore ceases to be valid.
As it will become clear in section 5
the truncation leading to (\ref{kappaflow}-\ref{e2flow})
is not a very good approximation for the region
of a first-order transition.
The low
value of $\lambda(\Lambda)/e^2(\Lambda)$ for the onset of
the second-order behaviour is therefore
likely to be an artefact of this approximation.
Nevertheless, the corresponding critical
$\lambda(\Lambda)/e^2(\Lambda)$ can be interpreted as a lower bound
for the change to a second-order behaviour.
One should notice that the form of trajectories shown
in Fig.~4 is reliable, and only the selection of the
``transition line'' from first- to second-order
behaviour is doubtful. The latter depends on global properties of the
flow equations outside the validity of the linear approximation
and the approximation leading to (\ref{kappaflow}-\ref{e2flow}).

There are other important shortcomings of the simple flow
equations (\ref{kappaflow}-\ref{e2flow})
which suggest that the phase diagram
Figs.~2-4 gives at best a qualitatively correct picture but no
quantitatively reliable details: first the
threshold function in the running of $e^2$
should be properly taken into account.
In contrast to the oversimplified picture of
Fig.~3 the second-order fixed point and the tricritical
point -- if they exist -- will not correspond
to the same fixed point value $e^2_\star$.
Second, the polynomial approximation of the average potential
$U_k(\rho)$ (4)
does not well describe a first-order transition.
For this situation one needs enough freedom so that
$U_k(\rho)$ can have more
than one local minimum. In order to better understand the
necessary modifications for a quantitative description
of the superconductor phase transition
we derive in the next sections the flow equations
from the exact non-perturbative
flow equation for the average action
of scalar QED~\cite{NPB427}.
The various approximations
made will be explained in detail.

\end{section}


\begin{section}{\names The Average Action for Gauge Theories}

An overview of the basic formalism
of the effective average action for the abelian Higgs
model will be presented in this section.
For a more complete discussion the
reader is referred to \cite{NPB427}.
In short, the effective average action
$\Gamma_k$ is a type of coarse grained free energy where all
fluctuations with momenta larger than an ``average scale''
$k$ are integrated out.
We will use here a formalism where
the physical properties of the system are encoded in an
explicitly gauge invariant effective action.
In order to guarantee gauge invariance
we follow closely the background field method~\cite{abbott}.
The spirit of this approach is in complete
analogy with the well known block spin approach of
thinning out degrees of freedom in a lattice~\cite{goldenfeld},
but here it is applied to the continuum.
The integration of high momentum modes, $q^2 > k^2$,
is equivalent to the formulation of  an effective theory
for ``average'' fields where the averaging extends
over a typical length scale $k^{-1}$.
The effective action of the field theory
(free energy) is determined
in the limit of vanishing average scale $k\rightarrow 0$,
{\it {i.e.}~} when all fluctuations are integrated out~\cite{CW}.
We implement an infrared cut-off
for the fluctuations by introducing
an additional term quadratic in the fields. Thus
the scale dependent generating functional $W_k$ is given by
\begin{eqnarray}
e^{W_k [ J,K_\mu; \bar{A}_\mu ] }&=&
  \int D \varphi D a_\mu \exp - \left\{
   S[\varphi,a_\mu; \bar{A}_\mu ] +
   \Delta_k S [\varphi,a_\mu; \bar{A}_\mu ]
   +\frac{1}{2 \alpha} \int d^3x (\partial_\mu a^\mu )^2\right. \nonumber \\
 & & \left.
 ~~~~~~~~~~~~~~\quad\quad\quad
 - \int d^3x \left( J^* \varphi + J \varphi^* + K_\mu
     a^\mu \right) \right\} \,\, .
\label{eins}
\end{eqnarray}
Here $\bar{A}_\mu$ is some arbitrary background field, while
$a_\mu$ and $\varphi$ denote the gauge and scalar
field fluctuations, respectively.
The gauge invariant action $S$ depends
only on  $\varphi$ and
$\bar{A}_\mu + a_{\mu}$.
We will later work in the Landau gauge,
$\alpha \rightarrow 0$, but for completeness we
will keep the gauge fixing parameter $\alpha$ arbitrary
throughout most of this section.
The $\Delta_k S$ term implements the smooth infrared cut-off.
We write $\Delta_k S = \Delta_k S^{(S)} + \Delta_k S^{(G)}$ with
\begin{eqnarray}
\Delta_k S^{(S)}&=&~~~~~\int d^3x ~\varphi^* R_k^{(S)}(-D^2[\bar{A}])
  \varphi
\label{zweia} \\
\Delta_k S^{(G)}&=&~~\frac{1}{2} \int d^3x ~a_\mu R_k^{(G)}(-\partial^2)
  a^\mu \nonumber \\
  & & + \frac{1}{2} \int d^3x ~(\partial_\nu a^\nu)
  \frac{\alpha^{-1} \tilde{R}_k^{(G)}(-\partial^2) -
        R_k^{(G)}(-\partial^2)}{(-\partial^2)}
  (\partial_\mu a^\mu)
\label{zweib}
\end{eqnarray}
and $D^2[\bar{A}]$ the covariant Laplacian
in the background field $\bar{A}$.
The choice of the functions $R_k(x)$ is relatively free within certain
constraints \cite{CW}. Here we use
\begin{eqnarray}
R_k(x) = Z_k \frac{x~e^{-x/k^2}}{1-e^{-x/k^2}}
\label{drei}
\end{eqnarray}
which obeys
\begin{eqnarray}
\lim_{k \rightarrow 0} R_k(x) &=& 0
\label{viera}, \;  \; \; \; \; \; \;
\lim_{k \rightarrow \infty} R_k(x)= \infty, \,\,\,\, \mbox{for
fixed $x$} \\
\lim_{x \rightarrow 0} R_k(x)&=& Z_k k^2.
\label{vierib}
\end{eqnarray}
The wave function renormalization may
differ for $R_k^{(S)}$,
$R_k^{(G)}$, and $\tilde{R}_k^{(G)}$
and will at the end be adapted to corresponding
constants in the kinetic term of $\Gamma_k$.
The effective average action is
obtained by a Legendre transform of $W_k$
\begin{eqnarray}
\Gamma_k [\bar{\varphi}, A_\mu , \bar{A}_\mu ]
&=& - W_k [J,K_\mu ; \bar{A}_\mu ] -
 \Delta_k S [\bar{\varphi}, A_\mu , \bar{A}_\mu ] \nonumber \\
& & \vspace{2cm}+\int d^3x \left( J^* \bar{\varphi} + J \bar{\varphi}^* +
        K_\mu (A^\mu-\bar{A}^\mu) \right)
\label{fuenf}
\end{eqnarray}
where we have introduced
$A_\mu = \bar{A}_\mu+\bar{a}_\mu$, and
$\bar{\varphi}$ and $\bar{a}_\mu$ are the usual
classical fields
(we omit the bar for $\varphi$ in the following).
The running of $\Gamma_k$ with
changing average scale follows
by varying $k$ for constant fields:
\begin{eqnarray}
\frac{\partial}{\partial t}
\Gamma_k [{\varphi}, A_\mu , \bar{A}_\mu ] =
\frac{1}{2} \mbox{Tr} \left\{ \left(
\Gamma_k^{(2)} [{\varphi}, A_\mu , \bar{A}_\mu ]
+ R_k[ \bar{A}_\mu ] \right)^{-1}
\frac{\partial R_k[ \bar{A}_\mu ] }{\partial t} \right\}
\,\, .
\label{sechs}
\end{eqnarray}
Here the trace accounts for the summation
over all the field degrees of freedom
and includes  an integration in
coordinate  or momentum space.
The inverse propagator
$\Gamma_k^{(2)}$ denotes
the matrix of second functional derivatives of $\Gamma_k$
with the background field $\bar{A}_{\mu}$ held fixed.
Being an exact flow equation eq.~(\ref{sechs}) has
the peculiarity of exhibiting
a one-loop structure. If we introduce the operator
$\tilde{\partial}_t$, which
expresses a $t$-derivative acting
only on $R_k$, one can formally write it as
\begin{eqnarray}
\frac{\partial}{\partial t}
\Gamma_k [ \varphi, A_\mu,  \bar{A}_\mu ] =
\frac{1}{2} \mbox{Tr} \;
\tilde{\partial}_t
\ln \left( \Gamma_k^{(2)} [\varphi, A_\mu , \bar{A}_\mu ]
+ R_k[ \bar{A}_\mu ] \right) \,\, .
\label{sieben}
\end{eqnarray}
The momentum integration implied by
eqs.~(\ref{sechs}) and (\ref{sieben}) is
both infrared- and ultraviolet-finite.
For momenta $q^2 \ll k^2$, $R_k$ regulates the trace in
the infrared, whereas for $q^2 \gg k^2$, $\partial_t R_k$ is exponentially
suppressed (see (\ref{drei}) with $x\equiv q^2$).
This property holds for any number
of dimensions.

The flow equation (\ref{sieben}) is gauge invariant under
background gauge transformations,
where $A_\mu$ and $\bar{A}_\mu$ are transformed simultaneously.
The conventional gauge invariant effective action~\cite{abbott}
is obtained from the solution of the flow equation
(\ref{sieben}) as
\begin{eqnarray}
\Gamma [\varphi, A_\mu ] =
\lim_{k \rightarrow 0} \bar\Gamma_k [
\varphi, A_\mu]
\label{neuna}
\end{eqnarray}
where
\begin{eqnarray}
\bar{\Gamma}_k [ \varphi, A_\mu ]&=&
\Gamma_k [ \varphi, A_\mu, \bar{A}_\mu = A_\mu ] \,\, .
\label{neunb}
\end{eqnarray}
However, although we are interested in $\bar{\Gamma}_k$
rather than in the full solution of (\ref{sieben}), we can not
merely identify $\bar{A}_\mu = A_\mu$ in the flow equation, since
$\bar\Gamma_k^{(2)}$ is different from $\Gamma_k^{(2)}$.
In general one has
\begin{eqnarray}
\Gamma_k [ \varphi, A_\mu, \bar{A}_\mu ] =
\bar{\Gamma}_k [ \varphi, A_\mu ] +
\Gamma_{\rm gf} [ A_\mu, \bar{A}_\mu ] +
\hat{\Gamma}_{{\rm gf},k} [ \varphi, A_\mu, \bar{A}_\mu ]
\label{zehn}
\end{eqnarray}
where
\begin{eqnarray}
\Gamma_ {\rm gf}[A_{\mu},\bar{A}_{\mu}] =
\frac{1}{2\alpha} \int d^3 x \left (
\partial_{\mu} (A^{\mu}-\bar{A}^{\mu}) \right )^2
\end{eqnarray}
and
$\hat\Gamma_{{\rm gf},k}$ plays the
role of a scale dependent correction to the gauge fixing term
which contains the usual counterterms.
The form of $\hat\Gamma_{{\rm gf},k}$ is completely fixed
by gauge invariance, and one can derive the following
functional differential
equation for the dependence of this generalized gauge fixing
functional on the background field
\cite{NPB427}:
\begin{eqnarray}
\frac{\delta}{\delta \bar{A}_\mu(x)}
\hat\Gamma_{{\rm gf},k}[\varphi,A,\bar{A}] =
\frac{1}{2}\mbox{Tr}\left\{\frac{\delta R_k[\bar{A}]}{\delta \bar{A}_\mu(x)}
\left(\Gamma_k^{(2)}[\varphi,A,\bar{A}]+R_k[\bar{A}]\right)^{-1}
\right\} \,\, .
\label{new_2}
\end{eqnarray}
This equation
encodes the fact that we are dealing with
a gauge theory, and we observe that
$\hat\Gamma_{{\rm gf},k}$ vanishes for
$k\rightarrow 0$ since $R_k \rightarrow 0$.
In addition, one can also derive generalized
Ward identities
\cite{Ellwanger}
in presence of an infrared cutoff.
With $\Gamma ' = \Gamma_k-\Gamma_{\rm{gf}}$
and remenbering that $R_k$ depends on $\bar A$
they read in momentum space
\begin{eqnarray}
\frac{\delta \Gamma '}{\delta \bar{\delta}}
\frac{\delta \Gamma '}{\delta \delta}
%
\frac{1}{\bar{e}}~q_\mu ~\frac{\delta \Gamma '}{\delta {A}_\mu (q)}
&\!\!\!+\!\!\!& \int\frac{d^d p}{(2\pi)^d}
\left[\frac{\delta \Gamma '}{\delta \varphi(p)}~
\varphi(p-q) - \frac{\delta \Gamma '}
{\delta \varphi^*(p)}~\varphi^*(p+q)\right]
= \nonumber \\
& & \hspace*{1cm}\int\frac{d^d p}{(2\pi)^d}[R^{(S)}_k(p)-R^{(S)}_k(p-q)]~
\left(\Gamma^{(2)}_k + R_k\right)^{-1}_{\varphi^*(p),\varphi(p-q)}
\label{WI}
\end{eqnarray}
The ``anomaly term'' on the right-hand side vanishes
for $k\rightarrow 0$ and the
usual Ward identities are recovered in this limit. There is
overlap in the information contained in (\ref{new_2}) and (\ref{WI}).
Both imply, for example, a vanishing photon mass for $k\rightarrow 0$
in the symmetric phase. The identities (\ref{new_2}) and (\ref{WI})
relate the couplings parametrizing $\hat\Gamma_{\rm{gf,k}}$ to
the ones parametrizing the gauge invariant kernel $\bar{\Gamma}_k$.
In principle, one has to include with each invariant in a given
truncation of $\bar{\Gamma}_k$ a whole set of terms in
$\hat\Gamma_{\rm{gf,k}}$ which are related by (\ref{new_2}) or (\ref{WI}).
On the other hand, if one uses a simple truncation for
$\hat\Gamma_{\rm{gf,k}}$, equations (\ref{new_2}) and (\ref{WI})
lend themselves
to non-trivial checks on the consistency of a given
truncation.

We choose in the present work the following
strategy \cite{NPB427}:
we redefine the average action depending on $\varphi$ and
$A_\mu$ by
\begin{eqnarray}
{\Gamma}_k [ \varphi, A_\mu ] =
\bar\Gamma_k [ \varphi, A_\mu ] + C_k [A_\mu ]
\label{dreizehn}
\end{eqnarray}
where the normalization functional $C_k [A_\mu ]$ is taken to be
gauge invariant
and obeys $C_0[A_{\mu}] = 0$.
The idea is to find an ansatz for $C_k$ that
absorbs the dominant effects coming from $\hat{\Gamma}_{{\rm gf},k}$
such that
\begin{eqnarray}
Q_k[ \varphi, A_\mu ] =
\hat{\Gamma}^{(2)}_{{\rm gf},k} [ \varphi, A_\mu,
\bar{A}_\mu = A_\mu ] -
C_k^{(2)} [ A_\mu ] \,\,
\label{vierzehn}
\end{eqnarray}
in the right-hand side of
(\ref{sieben}) is small. We use here
\begin{eqnarray}
C_k[A] = - \frac{1}{2}\mbox{Tr}\left\{\ln\left(R_k[A]+
{\cal{Z}}_k k^2+ {\cal{M}}^2_k\right)-
\ln\left({\cal{Z}}_k k^2+ {\cal{M}}^2_k\right)\right\}
\label{new_3}
\end{eqnarray}
where in the regime with spontaneous symmetry breaking
${\cal{Z}}_k=Z_{\varphi,k}$ and
${\cal{M}}^2_k = 2 U''_k(\rho_0) \rho_0$. For
a discussion of possible choices
for $C_k[A]$
we refer the reader to \cite{NPB427}.
By neglecting the contributions from
$Q_k$ we arrive at the following
(approximate) evolution equation for
$\Gamma_k [\varphi, A_\mu ]$
\begin{eqnarray}
\frac{\partial}{\partial t}
\Gamma_k [ \varphi, A_\mu ] =
\frac{1}{2} \mbox{Tr} \; \tilde{\partial}_t
\ln \left( \Gamma_k^{(2)} [\varphi, A_\mu ]
+ \Gamma_{\rm gf}^{(2)}
+ R_k[ A_\mu ] \right) +
\frac{\partial}{\partial t} C_k [ A_\mu ] \,\, .
\label{fuenfzehn}
\end{eqnarray}
This equation contains no more reference to the
background field $\bar{A}$ and
$\Gamma^{(2)}_k[\varphi,A_\mu]$ is the
standard second functional derivative of
$\Gamma_k[\varphi,A]$.
The gauge covariant flow equation (\ref{fuenfzehn}) is the starting point
of our investigation.
It has the form of a renormalization group improved
scale derivative of a one-loop formula
-- also $\frac{\partial}{\partial{t}}C_k[A]$ can be cast in this form --
and may be viewed as a differential form of the Schwinger-Dyson equation.
The flow equations for the $n$-point functions obtain directly
by differentiating eq.~(\ref{fuenfzehn}) with respect to the fields.
For the two-point function one finds a differential form of the associated gap
equation and this extends similarly to higher $n$-point functions.
The only approximation made as compared to the exact equation
(\ref{sechs}) is the neglecting of $Q_k$.
As stated above, equation (\ref{new_2}) can be used to check
the consistency of this strategy. We have computed the presumably most
important contribution to $Q_k$, namely the one from a mass term
$m^2_{\bar{a}}$
for the gauge invariant combination $(A_\mu-\bar{A}_\mu)$ in
$\hat\Gamma_{{\rm gf},k}$, and find this term small as
compared to the dynamically generated mass of the gauge field
in the phase with spontaneous symmetry breaking. At the
second-order fixed point discussed in section 6,
$m^2_{\bar{a}}$ turns out negative and small. It vanishes in
the limit $k\to 0$ as required, and its inclusion changes the
fixed point value of the gauge coupling by about $5\%$. The
changes in the scalar couplings are even smaller.

\end{section}


\begin{section}{\names Approximate Solutions}

As it stands, the evolution equation
(\ref{fuenfzehn}) is a functional differential equation
or, equivalently, an infinite system of
coupled non-linear differential equations
for the $n$-point
functions. Approximations
will be necessary for finding
solutions and
we consider here the two lowest terms
in a systematic derivative expansion \cite{{CW1},{critexp},{DE}}.
The fact that the expansion is systematic is, in practice,
of limited use unless the smallness of the neglected terms
can be established. It is reasonable to expect small
higher derivative contributions in the scalar sector.
Since the anomalous dimension $\eta_\varphi = - \partial_t \ln Z_\varphi$
turns out small, the inverse propagator is
essentially given by $q^2+m^2$. A similar reasoning is
not possible in the gauge sector where $\eta_{\scriptscriptstyle F}
= - \partial_t \ln Z_{\scriptscriptstyle F}$
approaches one for a second-order phase transition (see below).
Bearing in mind this limitation, we shall nevertheless
proceed in this way as a first attempt to understand the system.
Keeping only terms quadratic in the momentum
the most general form for the average action reads
\begin{eqnarray}
\Gamma_k[\varphi, A_\mu] &=& \int d^dx \left\{
   U_k(\rho) + \frac{1}{4} Z_{{\scriptscriptstyle F},k}(\rho) F_{\mu\nu}
   F^{\mu\nu}
\right. + Z_{\varphi,k}(\rho)\left( D_\mu[A] \varphi \right)^*
   \left( D^\mu[A] \varphi \right)\nonumber \\
&& \left.
   ~~~~~~~~~~  - \frac{1}{4} Y_{\varphi,k}(\rho) \rho\partial^2 \rho + \ldots
\right\} \,\,.
\label{sechzehn}
\end{eqnarray}
Here $U_k(\rho)$ is
the effective average potential,
which reduces to the free energy for a constant order parameter
in the limit $k \rightarrow 0$.
As usual,
$F_{\mu\nu} = \partial_\mu A_\nu - \partial_\nu A_\mu$ and
$D_\mu[A] = \partial_\mu + i \bar{e} A_\mu$ is the covariant derivative.
Note that because we choose to work with a
gauge invariant action the
number of invariants is substantially reduced.
The functions $Z(\rho)$ and $Y(\rho)$
can be interpreted as $\rho$-dependent
wave function renormalizations. They contain the information
about $n$-point functions with only two non-vanishing
external momenta, in the limit
of small momenta.
In the present work we will consider $Z_{\varphi,k}$
independent of $\rho$ and neglect $Y_{\varphi,k}$.
Up to the appropriate wave function renormalization
constants we therefore work with a standard covariant
kinetic term for the scalar field.
By expanding around appropriate configurations of the fields
one derives the
evolution
equations for the average potential and the different renormalization
factors \cite{NPB391}. In particular one
finds for the average potential
in this approximation

\begin{eqnarray}
\frac{\partial}{\partial t}\mbox{$U_k$} (\rho)& = &~~\frac{1}{4\pi^2}
\int\!dx~x^{\frac{1}{2}}
\tilde{\partial}_t \ln \left(\mbox{$Z_{{\scriptscriptstyle F},k}$}
P(x) + (\mbox{$Z_{{\scriptscriptstyle F},k}$}(\rho)
-\mbox{$Z_{{\scriptscriptstyle F},k}$}) x
+ 2 \mbox{$Z_{\varphi,k}$} \bar e^2 \rho
\right)  \nonumber \\
&&+\frac{1}{8\pi^2}\int\!dx~x^{\frac{1}{2}} \tilde{\partial}_t
\ln \left(\mbox{$Z_{\varphi,k}$} P(x) + \mbox{$U_k'$}(\rho) +
2 \rho \mbox{$U_k''$}(\rho) \right)  \nonumber\\
&&+ \frac{1}{8\pi^2} \int\!dx~x^{\frac{1}{2}} \tilde{\partial}_t
\ln \left(\mbox{$Z_{\varphi,k}$} P(x) + \mbox{$U_k'$}(\rho) \right) ,
\label{zweiundzwanzig}
\end{eqnarray}
where the inverse ``average propagator''
\begin{eqnarray}
P(x) = x + Z_k^{-1} R_k(x)
\label{propagator}
\end{eqnarray}
contains the infrared cut-off and primes denote partial
derivatives with respect to
$\rho$. We observe the appearance of
$M^2(k)$ and $m^2(k)$
in the first two terms in eq.~(\ref{zweiundzwanzig})
once $\rho$ is taken at the minimum of the potential at $\rho_0(k)$.
The last term is the contribution
of the ``would-be Goldstone boson''
present in our gauge $\alpha \rightarrow 0$.
More explicitly we have for $\tilde{\partial}_t$ acting on $P(x)$
\begin{eqnarray}
\tilde{\partial}_t P(x) =
\frac{\partial}{\partial t} P(x) +
(\frac{\partial}{\partial t} \ln Z_k)
(P(x)-x)
\label{vierundzwanzig}
\end{eqnarray}
and choose for $Z_k$ in $R_k^{(S)}$ and $R_k^{(G)}$
the $k$-dependent factors $Z_{\varphi,k}$
and $Z_{{\scriptscriptstyle F},k}$ respectively.
In the last term of the right-hand side we identify the anomalous
dimensions
\begin{eqnarray}
\eta_{\varphi}=-\frac{\partial\ln Z_{\varphi,k}}{\partial t}, \; \;
\label{etascalar}
\eta_{\scriptscriptstyle F}
=-\frac{\partial \ln Z_{{\scriptscriptstyle F},k}}{\partial t}~.
\label{etagauge}
\end{eqnarray}
In the present paper we will
neglect the term proportional to the anomalous dimension
in (\ref{vierundzwanzig}).

For the investigation of scaling solutions corresponding
to second- or weak first-order phase transitions it is convenient to
work with dimensionless, renormalized quantities. To this end we define
\begin{eqnarray}
u_k(\tilde{\rho})&=&k^{-3} U_k(\rho) \label{fuenfundzwanziga}\\[.5ex]
\tilde{\rho} &=&k^{-1} \mbox{$Z_{\varphi,k}$} \, \rho
\label{fuenfundzwanzigb}\\[.5ex]
e^2 &=& k^{-1} (\mbox{$Z_{{\scriptscriptstyle F},k}$})^{-1}
\bar{e}^2 \label{e2dless}\,\,
\end{eqnarray}
where $Z_{{\scriptscriptstyle F},k}$ is
defined as $Z_{{\scriptscriptstyle F},k}(\rho_0(k))$.
In terms of the above, the partial
differential equation (\ref{zweiundzwanzig}) can be rewritten as
\begin{eqnarray}
\partial_{t} u_k(\tilde{\rho})&=& -3\, u_k(\tilde{\rho}) +
(1+\eta_\varphi)\tilde{\rho}\ u'_k(\tilde{\rho}) +
4 v_3~\tilde\ell_0 (2e^2\tilde{\rho},
z_{\scriptscriptstyle F}(\tilde{\rho})) \nonumber \\[.5ex]
&&  +~2 v_3~\ell_0 (u'_k(\tilde{\rho})) +~
2 v_3~\ell_0 (u'_k(\tilde{\rho})+2\tilde{\rho} u''_k(\tilde{\rho})) \,\,
\label{flowpot}
\end{eqnarray}
where primes denote now partial derivatives with respect to
$\tilde{\rho}$ and $v^{-1}_3=8\pi^2$.
The threshold functions $\ell_0 (\omega)$
and $\tilde\ell_0 (\omega,z)$ are
given in the appendix.
The function $z_{\scriptscriptstyle F}(\tilde{\rho})$ is defined as
\begin{equation}
z_{\scriptscriptstyle F}(\tilde{\rho}) =
\frac{Z_{{\scriptscriptstyle F},k}(\tilde{\rho})}{Z_{{\scriptscriptstyle F},k}}
=\frac{Z_{{\scriptscriptstyle F},k}
(\tilde{\rho})}{Z_{{\scriptscriptstyle F},k}(\kappa)}
\end{equation}
and for the running of the dimensionless renormalized gauge coupling we have
\begin{eqnarray}
\partial_{t} e^2 = -e^2 + \eta_{\scriptscriptstyle F} \; e^2 \,\, .
\label{flowe2}
\end{eqnarray}
The system of equations (\ref{flowpot})
and (\ref{flowe2}) constitutes the basis for all our
investigations. We will give
its solutions in several different approximations in the
following sections. At this stage we emphasize that besides the
approximations already mentioned, eq.~(\ref{flowpot})
is exact up to contributions from
higher derivative terms appearing
in $\mbox{$\Gamma$}_k$. The main effect of these omitted
higher derivative terms is a modification of the momentum dependence of the
average propagator $P(x)$, eq.~(\ref{propagator}).
Since only a relatively small range
$x\approx k^2$ contributes to the momentum integrals we do not expect
qualitative changes from the modification of $P(x)$. On the other hand,
the neglected higher derivative terms may have a more important impact
on the running of the gauge coupling, eq.~(\ref{flowe2}).

We need expressions for the anomalous dimensions
$\eta_{\varphi}$ and $\eta_{\scriptscriptstyle F}$ and the
function $z_{\scriptscriptstyle F}({\tilde{\rho}})$.
For the computation of
$\eta_{\varphi}$ and $\eta_{\scriptscriptstyle F}$ we first neglect
the $\rho$-dependence
of $Z_{{\scriptscriptstyle F},k}$ by setting
$z_{\scriptscriptstyle F}=1$, whereby
$\tilde\ell_0 (\omega,z)=\ell_0 (\omega)$.
In this truncation of $\rho$-independent wave function renormalizations
the anomalous dimension of the scalar field reads

\begin{eqnarray}
\eta_{\varphi} = -\frac{32}{3}v_3 e^2
\ell_{1,1}(2 \lambda \kappa, 2 e^2 \kappa) +
\frac{16}{3} v_3 \lambda^2\kappa~m_{2,2}(2 \lambda \kappa , 0)
\label{scandim}
\end{eqnarray}

\noindent while its gauge field counterpart is

\begin{eqnarray}
\eta_{\scriptscriptstyle F} = \frac{4}{3}v_3\left(\ell_g(
2 \lambda \kappa, 2 e^2 \kappa) + \ell_c(2 \lambda \kappa)\right) e^2.
\label{ggandim}
\end{eqnarray}
Here the various threshold functions $\ell_{1,1}$,
$m_{2,2}$, $\ell_{g}$ and $\ell_c$ can be also found in
the appendix and we note that the part
$\sim \ell_c$ corresponds to the term
$\partial_{t}C_k[A]$ in eq.~(\ref{fuenfzehn}).
For a derivation of $\eta_{\varphi}$
and $\eta_{\scriptscriptstyle F}$ one has to specify the value of $\rho$
where $Z_{\varphi({\scriptscriptstyle F})}$
is defined and we have chosen $\rho_0(k)$,
{\it {i.e.}~} $Z_{\varphi,k} \equiv
Z_{\varphi,k} (\rho_0(k))$ and similar for
$Z_{\scriptscriptstyle F}$. We will see later that $|\eta_{\varphi}|$
turns out to be much smaller than one
such that the neglecting of the $\rho$-dependence
of $Z_\varphi$ can be justified.
The same holds for the
dependence of $Z_{\varphi}$ on the momentum
$q^2$ which would arise beyond the lowest order
in the derivative expansion. On the other hand,
$\eta_{\scriptscriptstyle F}$ will often be large. In particular,
if there exists a fixed point for $e^2$
this necessarily occurs for
$\eta_{{\scriptscriptstyle F}\star} = 1$ (cf.~eq.~(\ref{flowe2})).
As a consequence, for any such fixed point the anomalous
dimension of the gauge field exactly equals
its canonical dimension such that the renormalized gauge field
has scaling dimension one.
For $|\eta_{\scriptscriptstyle F}|$ around one or larger
there is no good reason to
expect that the $\rho$- and $q^2$-dependence of
$Z_{\scriptscriptstyle F}$ is a small effect.
We therefore believe that a computation
of the fixed point value $e^2_\star$ in the present truncation
is not quantitatively reliable.
Fortunately, this uncertainty in
$e^2_\star$ only moderately affects
the fixed point values of the other couplings,
provided $e^2_\star\kappa_\star \gg 1$
-- as it will turn out later to be the case for
the relevant fixed points.

Indeed, in the limit $e^2\rightarrow\infty$,
for any non-vanishing $\tilde\rho$,
the contribution from the gauge field fluctuations
$\sim\ell_{0} (2 e^2 \tilde{\rho})$ in eq.~(\ref{flowpot}) vanishes.
The only difference to the pure scalar theory
(the two component Heisenberg model)
is then the different value of the anomalous dimension
$\eta_{\varphi}$. Within our truncation we find
\begin{eqnarray}
\lim_{e^2 \kappa \rightarrow \infty} \eta_{\varphi} =
-\frac{16}{3}v_3\kappa^{-1}\ell_1(2\lambda\kappa)+
\frac{16}{3}v_3\lambda^2\kappa~
m_{2,2}(2\lambda\kappa,0)
\end{eqnarray}
whereas only the second term appears in the Heisenberg model.
We may consider the limit $e^2 \rightarrow \infty$
as an effective scalar theory with non-local
interactions. This is very similar to the case of $N$ charged scalar fields
where the limit
$e^2(\Lambda) \rightarrow \infty$,
$\lambda(\Lambda) \rightarrow \infty$
for fixed $\rho_{0}(\Lambda)$ corresponds to the
$CP^{N-1}$ model \cite{cpn}.
We have drawn in Fig.~7 the phase diagram
for the effective non-local scalar theory
for $e^2 \rightarrow \infty$ which follows from our flow equations.
It shows a second-order phase transition
with an associated fixed point. This is the analogue
of the Wilson-Fisher fixed point for $e^2=0$, but the
critical exponents are not the same due to the
difference in $\eta_{\varphi}$.
For large $e^2\kappa$ the flow equations
can be expanded in powers of
$(e^2\kappa)^{-1}$.
This extends to physical quantities like critical exponents.
Despite the uncertainty
in the exact value of $e^2_\star$
these exponents can therefore
be computed relatively accurately provided
$e^2_\star\kappa_\star$
turns out large.

Within our truncation the threshold function
$\ell_{g}$ does not vanish for $e^2\kappa
\rightarrow \infty$. There remains a contribution
depending on $\lambda\kappa$ and
$\ell_g+\ell_c$
even turns negative for small
$\lambda \kappa$
and large $e^2\kappa$.
For larger
$\lambda \kappa$ the sum
$\ell_g+\ell_c$  is positive also for
$e^2\kappa \rightarrow \infty$ and the
functions $\ell_g$ and $\ell_c$
vanish for $\lambda \kappa \rightarrow \infty$.
Only for positive
$\ell_g+\ell_c$ one has positive $\eta_{\scriptscriptstyle F}$
and a fixed point for $e^2$.
In this case very small values of $1/e^2$ tend to
increase with decreasing $k$.
The corresponding non-local operator appearing in the effective
non-local scalar theory for non-vanishing $1/e^2$ is therefore a
relevant perturbation. We have indicated on the critical line
in Fig.~7 the region of infrared instability of
this perturbation by diamonds,
In particular, the second-order fixed point
in Fig.~7 acquires a new unstable direction
for non-vanishing $1/e^2$. In our truncation
the trajectories starting on the
critical surface in the vicinity of the $e^2 \rightarrow \infty$
fixed point are attracted towards the second-order
fixed point with finite $e^2$.
On the other hand, in the approximation (\ref{ggandim})
there is also a region
(crosses on the critical line in Fig.~7)
where $\eta_{\scriptscriptstyle F}$ is negative and
$e^2$ diverges.
(In this case we stop the running
of $e^2$ at some very large value.)
A situation where $1/e^2$ asymptotically reaches
zero is not incompatible
with our overall picture. Nevertheless, the
appearance of a region with negative
$\eta_{\scriptscriptstyle F}$ may also be a pure artefact
of our approximation for
$Z_{{\scriptscriptstyle F},k}(\rho,q^2)$ by
the constant
$Z_{{\scriptscriptstyle F},k}(\rho_0(k),q^2=0)=Z_{{\scriptscriptstyle F},k}$.
As a last remark we point out that the linear
approximation for $\eta_{\scriptscriptstyle F}$ used in section 2
leads to fixed point values $2e^2_\star\kappa_\star
= 8.02$ for the second-order fixed point and
$2e^2_\star\kappa_\star
= 61.1$ for the tricritical point.
Both of these values are substantially larger than
one. Taking into account the threshold functions
decreases the value of
$\ell_g+\ell_c$ and therefore increases the value of
$e^2_\star$ and $e^2_\star \kappa_\star$. Consequently,
we predict the critical behaviour for the second-order fixed point to be
between the one of the linear approximation
for $\eta_{\scriptscriptstyle F}$ and the one of the $e^2 \rightarrow \infty$
fixed point of the non-local scalar model. The same situation
holds for the crossover behaviour
associated to the tricritical point.
The fixed point which corresponds to the
tricritical point in the linear approximation is for the
non-local scalar theory the gaussian fixed point
$\lambda_\star = 0$. We will see that in some of our truncations
the tricritical point actually
moves to $\lambda_\star = 0$,
$1/e^2_\star = 0$. Of course in this version
$e^2(k)$ has no fixed point since it
increases towards infinity.

We next turn to approximations
for the evolution equation
(\ref{flowpot}) for the scalar potential. An important problem
is our lack of knowledge about the form of
$z_{\scriptscriptstyle F}(\tilde{\rho})$. We will describe in this section two
approaches where
$z_{\scriptscriptstyle F}(\tilde{\rho})$ is put to one and
refer for a discussion of the influence
of the $\tilde{\rho}$ dependence
of the photon wave function renormalization
to section 7.
If the potential $u_k{({\tilde\rho}})$
has only one minimum, and the curvature around the minimum is
not too small, one would expect that
a steepest descent approximation around
this minimum should give a reliable answer.
In this case we propose
as our first method a polynomial
expansion
\begin{eqnarray}
u_k(\tilde\rho) = \sum_{n=2}^\infty \frac{u_n}{n!}
(\tilde{\rho}-\kappa)^n .
\label{potexp}
\end{eqnarray}
A truncation for relatively small $n$ should
already provide a quantitatively reasonable answer.
The lowest order in such a series of truncations
is described in section 2, with $\lambda = u_2$
and the running of $e^2$ further simplified in the linear
approximation. In section 6 we will give results
for  different truncations
including $n=4$ with running of $e^2$ according to
eqs.~(\ref{flowe2}) and (\ref{ggandim}).
This method should well describe quantitatively a possible
second-order fixed point if certain
consistency requirements are fulfilled.
As a first consistency condition
for this approximation we may require that
$\lambda_\star$ should not turn out very
small at the fixed point, since for a relatively
flat potential higher orders in the
sum (\ref{potexp}) may become important.
In addition, all deviations from this fixed
point except the one corresponding
to the relevant parameter $\kappa$ should
be sufficiently strongly
damped for $k\rightarrow 0$.
With these conditions only a relatively small
range of $\tilde{\rho}$ around $\kappa$ is
expected to be relevant for the behaviour
around the fixed point and the approximation
$Z_{{\scriptscriptstyle F},k}(\rho) = Z_{{\scriptscriptstyle F},k}$
seems justified.
The weakness of this method
is that extrema of $u_k(\tilde{\rho})$
which are different from the one at
$\tilde{\rho} = \kappa$ are difficult
to detect. This makes the method inappropriate for a
good quantitative description of a first-order
phase transition and presumably also
for the tricritical point.

Our second method applies to small values of the ratio
$\lambda /e^2$. Here the scalar fluctuations can be
neglected. We expect this method to work
quantitatively well for the first-order phase transition
in good type-I superconductors.
If the phase transition is sufficiently
strongly first-order there is not much running
of $e^2_{\scriptscriptstyle R}(k)$ and
similarly $Z_{\scriptscriptstyle F}(\rho)$
remains near one. The neglecting of
the $\rho$-dependence of
$Z_{\scriptscriptstyle F}(\rho)$ is then reasonable.
With these approximations the flow equations
can be solved explicitly and will be discussed in the
next section.
Unfortunately, in the interesting region
of large $\lambda /e^2$ this method
presumably fails, both due to the
inaccuracy of the approximation
$z_{\scriptscriptstyle F}(\tilde{\rho}) =1$ and because of the
neglecting of the scalar fluctuations.
Nevertheless, even within this
method we will see a transition
to a second-order behaviour.
This method overemphasizes the gauge field
fluctuations leading to a first-order transition
in comparison with the scalar fluctuations determining the
second-order behaviour. We therefore interpret
the critical ratio $(\lambda(\Lambda) /e^2(\Lambda))^{}_c$ separating,
within this method, the first- from the
second-order behaviour as an upper bound.
With the opposite arguments the critical
value $(\lambda(\Lambda) /e^2(\Lambda))^{}_c$
found with the method of a local polynomial
expansion may be viewed as a lower bound.

\end{section}


\begin{section}{\names The First-Order Transition in Type-I Superconductors}

In this section we neglect the contribution of the scalar fluctuations
to the flow equation for the average potential. Approximating in addition
$Z_{{\scriptscriptstyle F},k}(\rho)$ by a
constant $Z_{{\scriptscriptstyle F},k}$ and putting $Z_{\varphi,k}=1$ this
yields for the evolution equation (\ref{zweiundzwanzig})
\begin{eqnarray}
k{{\partial  U_k(\rho)}\over{\partial k}} = \frac{k^3}{2\pi^2}~
\ell_0(2e^2_{\scriptscriptstyle R}(k)\rho/k^2).
\label{an1new}
\end{eqnarray}
The only non-perturbative effect by which eq.~(\ref{an1new}) differs from
a one-loop approximation is the
running of $e^2_{\scriptscriptstyle R}(k)$. We will assume here
that the flow of $e^2(k)=e^2_{\scriptscriptstyle R}(k)/k$ has
an infrared fixed point
$e^2_\star$ and approximate the $\beta$-function of the dimensionless
gauge coupling
\begin{eqnarray}
\frac{\partial e^2}{\partial t} = -e^2 +\frac{e^4}{e^2_\star}.
\label{an2new}
\end{eqnarray}
In the linear approximation one would have
$e^2_\star=6\pi^2/\ell_{gc}=70.4$, where $\ell_{gc}=\ell_g(0,0)+\ell_c(0)$,
but here we keep $e^2_\star=6\pi^2/\ell_{\rm eff}$ as a free parameter.
The simplified flow equation (\ref{an2new}) is easily solved with
the initial value
$\bar e^2_\Lambda = e^2_{\scriptscriptstyle R}(\Lambda) = e^2(\Lambda)\Lambda$
\begin{eqnarray}
e^2_{\scriptscriptstyle R}(k)
=\frac{e^2_\star\bar e^2_\Lambda}{e^2_\star
+\bar e^2_\Lambda({1\over k}-{1\over \Lambda})}.
\label{an3new}
\end{eqnarray}
We observe a characteristic transition scale
that for $e^2(\Lambda) \ll e^2_\star$ is
\begin{eqnarray}
k_{\rm tr} = \frac{\bar e^2_\Lambda}{e^2_\star}.
\label{an4new}
\end{eqnarray}
For $k\gg k_{\rm tr}$ the change in $e^2_{\scriptscriptstyle R}$ is negligible
which corroborates the one-loop approximation, with
$e^2_{\scriptscriptstyle R}(k)\approx\bar e^2_\Lambda$ in eq.~(\ref{an1new}).
For a sufficiently strong first-order phase transition
one has $k_{\rm dis}\gg k_{\rm tr}$ and the one-loop result obtained
earlier \cite{HLM} is valid.
On the other hand, for $k_{\rm dis}\ll k_{\rm tr}$
the running of $e^2_{\scriptscriptstyle R}(k)$ turns
to a different regime where
$e^2_{\scriptscriptstyle R}(k)\approx e^2_\star k$.
This regime is relevant for the scaling
behaviour at a second-order phase transition.
We will explore in the following how the existence
of these two different regimes
affects the characteristics of the phase transition.

In the analytical investigation of this section
we approximate the threshold function

\begin{eqnarray}
\ell_0(\omega) = \frac{\ell_0}{1+\omega}
\label{an1}
\end{eqnarray}

\noindent where $\ell_0=\ell_0(0)$.
This shares the main properties of a generic threshold function
\cite{critexp} {\it {i.e.}~} for $\omega>-1$
it is finite and decreases monotonically,
has a pole at $\omega=-1$ and for large $\omega$
it goes as $\omega^{-1}$. In this
approximation we have to solve the differential equation

\begin{eqnarray}
{{\partial  U_k}\over{\partial k}} =
A~{{k^4}\over{k^2+2e^2_{\scriptscriptstyle R}(k)\rho}}
\label{an2}
\end{eqnarray}

\noindent where $A=\ell_0/2\pi^2=0.09$.
Though we have found the general analytical
solution for eq.~(\ref{an2}) with $e^2_{\scriptscriptstyle R}(k)$ given by
eq.~(\ref{an3new}) it is more instructive to consider the solution
with a further simplification for
$e^2_{\scriptscriptstyle R}(k)$. Details of the crossover from the
constant $e^2_{\scriptscriptstyle R}(k)\approx \bar e^2_\Lambda$
to the scaling behaviour
$e^2_{\scriptscriptstyle R}(k)\approx e^2_\star k$
are not very important. Hence we
discuss the case when $e^2_{\scriptscriptstyle R}$
exhibits the behaviour (for $k_{\rm tr} < \Lambda$)

\begin{equation} \label{an5}
e^2_{\scriptscriptstyle R} (k)=\left\{
\begin{array}{c@{\quad \mbox{for}\quad}l}
\bar e^2_\Lambda & k>k_{\rm tr} \\[2ex]
e^2_\star k & k<k_{\rm tr} .
\end{array} \right.
\end{equation}

\noindent In the first regime, constant $e^2_{\scriptscriptstyle R}$,
equation (\ref{an2}) becomes

\begin{eqnarray}
{{\partial  U_k}\over{\partial k}} = A~{{k^4}\over{k^2+2\bar e^2_\Lambda
\rho}}
\label{an7}
\end{eqnarray}

\noindent and its most general solution is

\begin{eqnarray}
U_k (\rho)=A~\left[{k^3\over3}-2\bar e^2_\Lambda \rho k
+ (2\bar e^2_\Lambda\rho)^{3\over2}~
\arctan\sqrt{{k^2\over{2\bar e^2_\Lambda\rho}}}~\right]
+G(\rho).
\label{an8}
\end{eqnarray}
The function $G(\rho)$ is independent of $k$. It contains the initial values
and will be specified by appropriate boundary conditions for $k=\Lambda$.
The most prominent
property of the solution (\ref{an8}) is the presence of a term
proportional to $\rho^{3\over2}$. A close relation between this term and the
$\varphi^3$ term one finds in conventional perturbation theory,
causing the phase transition to be first-order, will soon be
established. In the second regime where the dimensionless gauge
coupling is constant, equation (\ref{an2}) is now given by

\begin{eqnarray}
{{\partial  U_k}\over{\partial k}} = A~{{k^3}\over{k+2e^2_\star
\rho}}
\label{an9}
\end{eqnarray}

\noindent and has the general solution

\begin{eqnarray}
U_k (\rho)=A~\Bigl[{k^3\over3}-e^2_\star\rho k^2 - (2e^2_\star \rho)^2 k
- (2e^2_\star \rho)^{3}~
{\rm ln}\left(\frac{k+2e^2_\star \rho}{2e^2_\star \rho}\right)\Bigl]
+F(\rho).
\label{an10}
\end{eqnarray}
Again, $F(\rho)$ is to be fixed by the boundary
conditions and (\ref{an10}) is valid for $0<k<k_{\rm tr}$.

The computation of the effective potential is completed by choosing
the boundary conditions and taking the $k\to0$ limit. For $k=\Lambda$
we consider

\begin{eqnarray}
U_{\Lambda}(\rho)=-\bar\mu^2_{\Lambda} \rho +
{1\over2}\bar\lambda_{\Lambda}\rho^2
\label{an11}
\end{eqnarray}

\noindent where $-\bar\mu^2_{\Lambda}$
and $\bar\lambda_{\Lambda}$ are respectively the mass
square and the scalar self-coupling at $k=\Lambda$.
Whilst $G(\rho)$ is determined by (\ref{an11}), $F(\rho)$ is
fixed by equating the right-hand sides of (\ref{an8}) and (\ref{an10})
at $k=k_{\rm tr}$. One then finds for $k>k_{\rm tr}$

\begin{eqnarray}
U_k (\rho)\!&=&\!-\bar\mu^2_{\Lambda} \rho
+ {1\over2}\bar\lambda_{\Lambda}\rho^2
+{A\over3}~(k^3-\Lambda^3)-2A~\bar e^2_\Lambda\rho(k-\Lambda) \nonumber\\
& &\! +A~(2 \bar e^2_\Lambda\rho)^{3\over2}
\left(\arctan \sqrt{k^2\over{2\bar e^2_\Lambda\rho}}
-\arctan \sqrt{\Lambda^2\over{2\bar e^2_\Lambda\rho}}~\right)
\label{an11b}
\end{eqnarray}
and for $k<k_{\rm tr}$

\begin{eqnarray}
U_k (\rho)\!&=&\!-\bar\mu^2_{\Lambda} \rho
+ {1\over2}\bar\lambda_{\Lambda}\rho^2
+{A\over3}~(k^3-\Lambda^3)
-A~e^2_\star\rho (k^2+k^2_{\rm tr}-2\Lambda k_{\rm tr} ) \nonumber\\
& &\! +A~(2 e^2_\star\rho)^2_{} (k-k_{\rm tr} ) - A~(2e^2_\star\rho )^3
\ln\Big({{2e^2_\star\rho + k}\over{2e^2_\star\rho
+ k_{\rm tr}}}\Big) \nonumber\\
& &\! +A~(2 e^2_\star\rho k_{\rm tr})^{3\over2}
\left(\arctan \sqrt{k_{\rm tr}\over{2e^2_\star\rho}}
-\arctan \sqrt{\Lambda^2\over{2e^2_\star\rho k_{\rm tr}}}~\right).
\label{an12}
\end{eqnarray}
Finally in the limit $k\to0$ one obtains (up to an irrelevant constant)

\begin{eqnarray}
U(\rho)\!&=&\!-(\bar\mu^2_{\Lambda}+A~e^2_\star(k^2_{\rm tr}
- 2\Lambda k_{\rm tr}))\rho +
{1\over2}(\bar\lambda_{\Lambda}-8A~e^4_\star k_{\rm tr})\rho^2
\nonumber\\
& &\! +A~(2 e^2_\star\rho k_{\rm tr})^{3\over2}
\left(\arctan \sqrt{k_{\rm tr}\over{2e^2_\star\rho}}
-\arctan \sqrt{\Lambda^2\over{2e^2_\star\rho k_{\rm tr}}}~\right)
\nonumber\\
& &\!+ A~(2e^2_\star\rho )^3
\ln \left(1+{k_{\rm tr}\over{2e^2_\star\rho}}\right).
\label{an13}
\end{eqnarray}
For a given $e^2_\star$ the form of the potential mainly depends on two
ratios of mass scales, namely
\begin{eqnarray}
\frac{k_{\rm tr}}{\Lambda}=
\frac{\bar e^2_\Lambda}{e^2_\star \Lambda}=\frac{16\pi\alpha}{e^2_\star}
\label{an13b}
\end{eqnarray}
-- the last identity holds for standard superconductors -- and

\begin{eqnarray}
\frac{\mu^2_{\rm eff}}{\Lambda^2}&=&\frac{\bar\mu^2_\Lambda}{\Lambda^2}-
2A~e^2_\star\frac{k_{\rm tr}}{\Lambda}+
A~e^2_\star\frac{k^2_{\rm tr}}{\Lambda^2} \nonumber\\
\nonumber\\
&=&\frac{\bar\mu^2_\Lambda}{\Lambda^2}-
2A~\frac{\bar e^2_\Lambda}{\Lambda}+
A~\bar e^2_\Lambda\frac{k_{\rm tr}}{\Lambda^2}~.
\label{an13c}
\end{eqnarray}
It is instructive to consider the
limits $k_{\rm tr}\to0$ and $k_{\rm tr}\to \Lambda$
for fixed $\bar e^2_\Lambda$. For the former, the logarithm can be expanded
and we obtain the perturbative one-loop result

\begin{eqnarray}
U (\rho)\!&=&\!-(\bar\mu^2_{\Lambda}-2A~\bar e^2_\Lambda \Lambda)\rho
+ {1\over2}\bar\lambda_{\Lambda}\rho^2 - A~(2 \bar e^2_\Lambda\rho)^{3\over2}
\arctan \sqrt{\Lambda^2\over{2\bar e^2_\Lambda\rho}}\,.
\label{an13d}
\end{eqnarray}
For $\mu^2_{\rm eff}=0$ the leading term for small $\rho$ is
$-\sqrt{2} \pi A~\bar e^3_\Lambda \rho^{\frac{3}{2}}$.
This is responsible for turning the phase transition first-order in a
type-I superconductor~\cite{HLM} in contradistinction to the second-order one
predicted by mean field theory.

In the other limit, $k_{\rm tr} \to \Lambda$,
the second last term on the right-hand side
of (\ref{an13}) vanishes such that

\begin{eqnarray}
U(\rho)\!&=&\!-(\bar\mu^2_\Lambda-A~e^2_\star \Lambda^2)\rho
+ {1\over2}(\bar\lambda_\Lambda-8A~e^4_\star \Lambda)\rho^2
+8A~e^6_\star\rho^3\ln \left(1+\frac{\Lambda}{2e^2_\star\rho}\right).
\label{an14}
\end{eqnarray}
For $\mu^2_{\rm eff}=0$ the leading term for small $\rho$ is now the
term $\sim\rho^2$. We distinguish two cases: for
$\bar\lambda_\Lambda/\bar e^2_\Lambda < 8A~e^2_\star$,
the phase transition from varying
$\mu^2_{\rm eff}$ remains first-order also in this limiting case; and
for $\bar\lambda_\Lambda/\bar e^2_\Lambda > 8A~e^2_\star$, we observe now a
second-order transition. There is a critical ratio

\begin{eqnarray}
\left(\frac{\bar\lambda(\Lambda)}{\bar e^2(\Lambda)}\right)_{\rm crit} =
8A~e^2_\star  \label{an15}
\end{eqnarray}
which separates the first- from the second-order behaviour.
Of course, the limit $k_{\rm tr}\to \Lambda$ corresponds to values of
$\bar e^2_\Lambda$ much larger than those for standard superconductors
(cf.~eq.~(\ref{an13b})). Also $8Ae^2_\star$ is substantially larger than
one, such that for $\bar\lambda_\Lambda$ as large as the critical value
(\ref{an15}) the neglecting of scalar fluctuations is no longer
justified.

Despite this, it is interesting to understand the solution (\ref{an13})
also for arbitrary values of $k_{\rm tr}$. It is convenient to switch to
dimensionless variables
\begin{eqnarray}
r=\frac{2e^2_\star\rho}{k_{\rm tr}}~~~~~~
\sigma=\frac{\mu^2_ {\rm eff}}{2e^2_\star k^2_{\rm tr}}~~~~~~
l=\frac{\bar\lambda_\Lambda}{4e^4_\star k_{\rm tr}}~~~~~~
\hat u=\frac{U}{k^3_{\rm tr}}
\end{eqnarray}
where
\begin{eqnarray}
\hat u(r)&=&-\sigma r + \frac{l-2A}{2} r^2
+ Ar^{\frac{3}{2}}\left(\arctan\sqrt\frac{1}{r}
-\arctan\sqrt\frac{\Lambda^2}{k^2_{\rm tr}r}~\right)+
Ar^3\ln\left(\frac{1+r}{r}\right).
\label{dimpot}
\end{eqnarray}
The extrema of $\hat u(r)$
obey  $\frac{\partial \tilde u}{\partial r}=\sigma$ where
$\tilde u (r) = \hat u (r) + \sigma r$ and
\begin{eqnarray}
\frac{\partial \tilde u}{\partial r}&=&(l-2A)~r
+ \frac{3}{2} Ar^{\mbox{\small{$\frac{1}{2}$}}}\left(\arctan\sqrt\frac{1}{r}
-\arctan\sqrt\frac{\Lambda^2}{k^2_{\rm tr}r}~\right)
\nonumber \\ \label{g74}
&&-\frac{A}{2}~\frac{r}{1+r} + \frac{A}{2}~\frac{k_{\rm tr}}{\Lambda}
\frac{r}{1+\frac{k^2_{\rm tr}}{\Lambda^2}~r}
+3A~r^2\ln\left(\frac{1+r}{r}\right)-A\frac{r^2}{1+r}~.
\label{eqnstate}
\end{eqnarray}

For an investigation of the behaviour
of $\frac{\partial \tilde u}{\partial r}$ we first
expand for small values of $r$
\begin{eqnarray}
\frac{\partial \tilde u}{\partial r}&=& \left(l-4A+2A~
\frac{k_{\rm tr}}{\Lambda}\right)r
-A\left( 3\ln r
+\frac{k^3_{\rm tr}}{\Lambda^3}\right)r^2
+\frac{4}{5}A\left(4+\frac{k^5_{\rm tr}}{\Lambda^5}\right)~r^3
+{\cal O}(r^4)~.
\label{g75}
\end{eqnarray}
One can check that $\frac{\partial \tilde u}{\partial r}$
increases with $r$ for large values of $r$.
We show the form of
$\frac{\partial \tilde u}{\partial r}$
for $l<(4-2k_{\rm tr}/\Lambda)A$ in Fig.~8a
and similarly for
$l>(4-2k_{\rm tr}/\Lambda)A$ in Fig.~8b.
The critical value  $l_c=(4-2k_{\rm tr}/\Lambda)A$ corresponds to
\begin{equation}
\left(\frac{\bar \lambda(\Lambda)}{\bar e^2(\Lambda)}\right)_c
=8Ae^2_\star\left(2-\frac{k_{\rm tr}}{\Lambda}\right)~.
\label{g76}
\end{equation}
For $k_{\rm tr}=\Lambda$ we recover (\ref{an15}) whereas
for $k_{\rm tr}/\Lambda\rightarrow 0$ there is an
additional factor of two.
Note that in Figs.~8 the axis
$\frac{\partial \tilde u}{\partial r}=\sigma$ can be shifted
parallel to itself as the value of $\sigma$ is varied.
Hence depending on $\sigma$ this axis will intersect
the curve at different points or it might not intersect it at all.
In particular, for $l<l_c$ there is the possibility of
intersecting the curve at two distinct points as illustrated in Fig.~8a.
These correspond to a local maximum, the one near the origin, and to a
local minimum, the one far to the right. When the axis does not
intersect the curve the superconductor is in the normal phase and the minimum
of $u$ is at the origin. As the
axis moves upwards it will touch the curve at a point
that subsequently splits into the two local extrema mentioned above.
Eventually only the minimum for $r\neq0$ will pervade for positive $\sigma$.
This sequence describes the
transformations one expects to take place on the effective potential
as the system undergoes a first-order phase transition. In fact, we have
$\sigma \sim (T_c-T-\Delta T)$ with
$\Delta T\approx T_c-T_{\rm un}$ and $T_{\rm un}$
the temperature where the symmetric phase becomes classically unstable.
An analogous discussion for $l>l_c$ leads to the idea that in
this case the system undergoes a second-order phase transition.

If the critical value eq.~(\ref{g76}) separates
the region of a first-order transition
from the one of a second-order transition we expect for small $k$ a
scaling solution of $u_k(\rho)$, if $\bar\mu^2_\Lambda$ is tuned to be on the
critical surface. This scaling solution should describe the tricritical
fixed point. Writing the potential (\ref{an12}) in terms of the dimensionless
variables introduced in the last section
(\ref{fuenfundzwanziga}-\ref{e2dless}) one finds for
$k \ll k_{\rm tr}$ (neglecting the constant)
\begin{eqnarray}
u_k(\tilde\rho)&=&-
(\frac{\mu^2_{\rm eff}}{k^2}
+A e^2_\star) \tilde\rho
+\frac{1}{2 k} \Bigl[
\bar\lambda_\Lambda + 8 A e^4_\star
 - 8 A~e^4_\star k_{\rm tr}
\left(2 -
\frac{k_{\rm tr}}{\Lambda} \right)\Bigl] \tilde\rho^2\nonumber\\
&&-8A~e^6_\star\Bigl[\ln\frac{k}{k_{\rm tr}}+\ln(1+2e^2_\star\tilde\rho)
-\frac{1}{3}\left(1-\frac{k^3_{\rm tr}}{\Lambda^3}\right)\Bigl]\tilde\rho^3
+{\cal O}\left(\frac{k}{k_{\rm tr}}\right)~.
\label{g77}
\end{eqnarray}
For $\mu^2_ {\rm eff}=0$ and $\bar\lambda_\Lambda$
given by eq.~(\ref{g76}) we find that
$u_k(\tilde\rho)$ becomes indeed almost independent of $k$, except for the
logarithmic running of the $\tilde\rho^3$ term. Defining
$\lambda(k)=u_2(\kappa)$ as before this leads to a logarithmic increase
of $\lambda$ for $k\rightarrow 0$. We would expect that the inclusion of
scalar fluctuations would stop this logarithmic running for the true
tricritical behaviour.

In summary, we have found indications for a transition to a second-order
behaviour for large $\bar\lambda(\Lambda)/\bar e^2(\Lambda)$ even within an
approximation where the contribution of the scalar fluctuations to the
free energy is retained only for the running of $e^2$.
The crude treatment of the running of $e^2_{\scriptscriptstyle R}(k)$
is not crucial in this respect. The inclusion of scalar fluctuations
for the potential in general
tends to stabilize a second-order behaviour. Consequently, we interpret
the critical value eq.~(\ref{g76}) as an upper bound for the true separation
between first- and second-order behaviour.

The analytical discussion of this section
is thought to be quantitatively reliable for small enough
$\bar\lambda(\Lambda)/\bar e^2(\Lambda)$,
say $\bar\lambda(\Lambda)<\mbox{\small{$\frac{1}{3}$}}\bar e^2(\Lambda)$.
Up to small corrections from the scalar
contributions and some modifications from
the $\rho$-dependence of $z_{\scriptscriptstyle F}$
(which should be minor for small values of
$e^2(\Lambda)$ in standard superconductors) we may therefore use the free
energy, eq.~(\ref{an13}), for a computation of the discontinuity of the
order parameter, the latent heat, etc.
Typical mass scales are proportional to
$k_ {\rm dis}$ (\ref{discontinuity}).

In fact, for the small values of $l$ for
which the discussion of this section is
quantitatively reliable the range in $r$ relevant
for the phase transition in
standard superconductors obeys (for $\Lambda$ near $T$)
\begin{eqnarray}
\frac{k^2_{\rm tr}}{\Lambda^2}r~\ll~1~\ll~r
\label{A}
\end{eqnarray}
\begin{eqnarray}
\frac{k^2_{\rm tr}}{\Lambda^2}=
\left(\frac{\bar e^2_\Lambda}{e^2_\star \Lambda}\right)^2=
\left(\frac{16\pi\alpha T}{e^2_\star \Lambda}\right)^2~\ll~1~.
\label{B}
\end{eqnarray}
We can therefore expand the potential (\ref{dimpot})
\begin{eqnarray}
\hat u(r) = - \tilde\sigma r -\frac{\pi}{2}A~r^{\frac{3}{2}}+
{1\over2} \tilde l r^2
\label{C}
\end{eqnarray}
where
\begin{equation}
\tilde\sigma=\frac{e^2_\star \bar\mu^2_\Lambda}{2 \bar e^4_\Lambda}-
A~\frac{\Lambda}{k_{\rm tr}}~~~~~~
\tilde l=l + \frac{2A~k_{\rm tr}}{\Lambda}~.
\label{D}
\end{equation}
For large enough $\tilde\sigma$ a local minimum occurs for $r>0$ at
\begin{eqnarray}
r^{1\over2}_{\rm min} = \frac{3\pi A}{8\tilde l} +
\left(\frac{9\pi^2 A^2}{64\tilde l^2}+
\frac{\tilde\sigma}{\tilde l}\right)^{1\over2}~.
\label{E}
\end{eqnarray}
The value $\tilde\sigma_c$ corresponding to the critical
temperature for the phase
transition is defined by $\hat u(r_{\rm min})=\hat u(0)=0$ which yields
\begin{eqnarray}
\tilde\sigma_c=-\frac{\pi A}{4}r^{1\over2}_{\rm min}=
-\frac{\pi^2 A^2}{8\tilde l}~.
\label{F}
\end{eqnarray}
At the critical temperature we therefore find for
the discontinuity of the order
parameter
\begin{eqnarray}
r_{\rm min}&=&\left(\frac{\pi A}{2 \tilde l}\right)^2 \nonumber\\
\nonumber\\
\rho_{\rm min}&=&2\pi^2 A^2 \frac{\bar e^6_\Lambda}{\bar\lambda^2_\Lambda}
\left(1+\frac{8A\bar e^4_\Lambda}{\bar\lambda_\Lambda \Lambda}\right)^{-2}
\nonumber\\
&=&\frac{\pi^2}{8} \left(\frac{\ell_0}{\ell_2}\right)^2
\frac{k^2_{\rm dis}}{\bar e^2_\Lambda} \left(1+\frac{2\ell_0}{\ell_2}
\frac{k_{\rm dis}}{\Lambda}\right)^{-2}.
\label{G}
\end{eqnarray}
We observe that in this approximation the
running of $e^2_{\scriptscriptstyle R}$ has no effect.
Quantitatively, the scale dependence of the gauge coupling will only appear in
higher orders in an expansion of $\hat u(r)$ in powers of $1/r$. The photon
mass term at the critical temperature obeys in the approximation (\ref{C})
\begin{eqnarray}
M=\frac{\pi \ell_0}{2\ell_2} \frac{e(k_{\rm dis})}{e(\Lambda)}
\left(1+\frac{2\ell_0}{\ell_2}
\frac{k_{\rm dis}}{\Lambda}\right)^{-1} k_{\rm dis}~.
\label{H}
\end{eqnarray}

For large values of $\bar\lambda(\Lambda)$ the
scalar fluctuations become important.
Nevertheless, for a first-order transition the ratio $\lambda(k)/e^2(k)$ will
always become small for sufficiently small
$k$ (cf.~eq.~(\ref{betaratio}) and Fig.~6).
We therefore propose the following strategy
for a quantitative computation of the first-order phase transition for
$\bar\lambda(\Lambda)/\bar e^2(\Lambda)
{\:\hbox to -0.2pt{\lower2.5pt\hbox{$\sim$}\hss}
           {\raise3pt\hbox{$>$}}\:} 1$: for scales $k$ where
$\lambda(k)/e^2(k)$ remains larger than a small constant (say $1/5$) we use
the running of
couplings as described by the polynomial expansion around the minimum of
$U_k$ (section~4). At the scale $\bar k$
where $\lambda(\bar k)/e^2(\bar k)=1/5$
we switch to the description of this section.
In fact, the approach used in this
section remains valid if we replace $\Lambda\rightarrow \bar k$,
$\bar e^2(\Lambda)\rightarrow e^2(\bar k)\bar k$,
$\bar\lambda(\Lambda)\rightarrow \lambda(\bar k)\bar k$, etc.
The details of the phase transition which involve scales $k<\bar k$
can then be extracted from the modified version of eq.~(\ref{an13}). We note
that $k_{\rm tr}/\bar k$ may now be substantially larger than for good type-I
superconductors where $\bar k=\Lambda$. Especially, for $e^2(\bar k)$
near $e^2_\star$ the equation of state should be close to
the generalization of eq.~(\ref{eqnstate}) for $e^2_{\scriptscriptstyle R}(k)$
given by (\ref{an3new}).

\end{section}


\begin{section}
{\names The Second-Order Transition in Strongly Type-II Superconductors}

In this section we discuss a numerical study
of the flow equations of the abelian
Higgs model, using the polynomial expansion that we have
developed in section~4. As we have seen in section 5,
for a small ratio $\lambda/e^2$
the system undergoes a first-order
phase transition. The effects of scalar fluctuations are small in
this case and may be neglected.
On the other hand, in the absence of
gauge fluctuations ($e^2=0)$ the
phase transition was shown to be second-order~\cite{critexp}.
Therefore one would expect that there will be
some sets of parameters $\lambda$ and $e^2$ for which the system is
passing from a first- to a second-order behaviour.
In the last section we gave some analytic solutions
which establish the region of a
first-order transition. Identifying the parameter region
for a  second-order transition analytically is not
easy, therefore in this part we have employed
numerical methods for a solution of the flow equations.
For the running of $e^2$ we insert (\ref{ggandim})
into the evolution equation (\ref{flowe2})
and the scalar anomalous dimension is specified by eq.~(\ref{scandim}).
In lowest order the running of $\kappa$ and $\lambda$ follows
eqs.~(\ref{kappaflow}) and (\ref{lambdaflow}).
Going to higher terms in the expansion of $u_k(\tilde\rho)$ in powers of
$(\tilde\rho-\kappa)$ one has also to include here contributions from $u_3$
and $u_4$. The form of these contributions, as well as the flow equations
for $u_3$ and $u_4$, follows by taking appropriate partial derivatives of
eq.~(\ref{flowpot}) with respect to $\tilde\rho$.
Throughout this section we use
$z_{\scriptscriptstyle F}(\tilde\rho)=1$.

At the critical temperature of a second-order
phase transition the theory has a fixed point, that is
the dimensionless parameters $\kappa$, $\lambda$ and $e^2$
do not depend on $k$ -- while the dimensionful quantities vanish with
appropriate powers of the scale.
Then, the first step in constructing a phase diagram for the
abelian Higgs model is to explore the
fixed point structure of the theory.
Finding the fixed point is not completely straightforward since we do not
dispose of an analytical form of the $\beta$-functions. We have employed
the following method: we set the initial conditions for the couplings
and the minimum of the potential at a short-distance scale $k = \Lambda$
and follow the $k$-dependence of the flow equations down to
$k=0$. This solution gives the vacuum expectation value
$\rho_0$ and the renormalized gauge
and quartic couplings at zero momentum. At the second-order fixed
point the system of flow equations is infrared stable in all
directions, except one which corresponds to the relevant
parameter $\sim (T_c - T)$ and may be associated with $\kappa$.
This indicates that if we change the sign of the flow of $\kappa$ and do
the renormalization group running, the system of flow equations will
run to the fixed point, provided the initial
values $\lambda(\Lambda)$ and $e^2(\Lambda)$
are within its domain of attraction. Since in order to
establish the existence of fixed points we are interested
only in the zeros of the $\beta$-functions,
this change of sign does not affect our findings in any way.

In order to check whether our results are stable in different truncations,
we have studied two distinct approximations of the potential as a local
polynomial in $(\tilde{\rho}-\kappa)$ up to $(\tilde{\rho}-\kappa)^2$
and $(\tilde{\rho}-\kappa)^4$, which we denote by $\varphi^4$-
and $\varphi^8$-approximation, respectively. For the first approximation
we have therefore
to solve a system of differential equations for the three functions
$\kappa(t)$, $\lambda(t)$ and $e^2(t)$,
whereas for the $\varphi^8$-approximation the
system is enlarged by the two additional functions $u_3(t)$ and $u_4(t)$.
We find that at the second-order fixed point
\begin{eqnarray}
&\varphi^4:&  \; \; \; \; \; e^2_\star =  634, \; \; \;
\kappa_\star =  0.045, \; \; \;
\lambda_\star =  25.6  \label{fpphi4} \\[1ex]
&\varphi^8:&  \; \; \; \; \; e^2_\star =  912,
\; \; \; \kappa_\star =  0.066, \; \; \;
\lambda_\star =  13.3 , \; \; \;
u_{3\star} = 203, \; \; \; u_{4\star} = 2345. \; \; \;
\label{fpphi8}
\end{eqnarray}
We see that in the second case $e^2_\star$ changes substantially.
Also $\kappa_\star$ and $\lambda_\star$ change substantially by going from the
$\varphi^4$- to the $\varphi^8$-approximation.
The main effect is due to the inclusion of the
$\varphi^6$-coupling which is dimensionless in $d=3$.
However, experience from the pure scalar
theory~\cite{critexp} suggests that the inclusion of additional terms
({\it {e.g.}~} $\sim\varphi^{10}$) will
give no further important modifications.

Once we have found the second-order fixed point, it is
easy to draw the phase diagram around it by plotting the trajectories
attracted to the fixed point. As we can see in Fig.~2
there exists a critical line which separates the broken from the
symmetric phase. The crossing between the two regimes is continuous,
as it is to be expected for a second-order phase transition.
For points below this critical line, the minimum $\kappa$ runs
to $0$ at some finite scale, indicating that we are in the regime
where the symmetry is restored. On the other hand, above this line
the coupling $\kappa$ scales as $k^{-1}$ and symmetry breaking occurs
for a value $\rho_0(k=0) = \rho_0 > 0$,
where $\rho_0 = Z^{-1}_\varphi(k)\kappa k$.

To confirm our results, we have also looked at the
critical exponents of the theory. These numbers describe the behaviour
of a system near the critical temperature of
a second-order phase transition, by specifying the long range correlations
of physical quantities as we approach a zero mass theory.
All relevant quantities may be calculated once the temperature
dependence of the correlation length and the behaviour of the connected
two-point function at $T_{c}$ (described by the $\nu$ and $\eta$
exponent respectively) are known. For this reason, we focused
on the behaviour of these two critical indices. For the second-order
fixed point we find
\begin{eqnarray}
&\varphi^4:&  \; \; \; \; \; \eta_\star =
-0.134, \; \; \; \; \nu_\star = 0.532\\[1ex]
&\varphi^8:&  \; \; \; \; \; \eta_\star =
-0.170, \; \; \; \; \nu_\star =  0.583.
\end{eqnarray}
These numbers indicate that in both truncations,
the critical exponents belong in the physical
region $\eta > 2-d$, and $\nu > 0$,
thus pointing clearly towards a second-order phase transition.
Moreover, our predictions are in satisfactory agreement with the measured
exponents for the phase transition of the nematic-to-smectic-A liquid
crystals within the experimental uncertainties~\cite{{nemdata},{newdata}}.
This transition is speculated to be in
the same universality class as superconductors.

On the surface which separates the
symmetric phase (normal conductor) from the one with spontaneous symmetry
breaking (superconductor) there also exists a second critical line
(hypersurface) which limits the region of attraction
towards the second-order fixed point (cf.~Fig.~3). It separates the region in
coupling constant space where the transition is second-order from the one
of a first-order transition. On this line the couplings flow towards
a different fixed point, the tricritical fixed point.
The tricritical point has two unstable directions
in $\kappa$ and $\lambda$, thus in order to search for
it we have inverted the signs of the relevant $\beta$-functions,
in complete analogy to the procedure we followed in order to discover
the second-order fixed point. However we find that the gauge coupling with
the full equations,~(\ref{flowe2}) and (\ref{ggandim}),
runs to infinity in the interesting region.
As we have discussed in section 4 this feature is likely to be an artefact
of the  approximations involved. Looking at the threshold function we see
that $\eta_{\scriptscriptstyle F}$ turns
negative in the parameter region depicted in
Fig.~9. We expect that once the parameters enter in that region
the evolution equation for $\beta_{e^2}$ will drive $e^2$ to infinity.
In this limit the fixed points of the theory behave like the fixed points
of a non-local scalar model. The tricritical fixed point is then given by
\begin{eqnarray}
&\varphi^4:&  \; \; \; \; \; e^2_{\rm tric} = \infty, \;
\kappa_{\rm tric} = 0.15, \;
\lambda_{\rm tric} = 0 \\[1ex]
&\varphi^8:&  \; \; \; \; \; e^2_{\rm tric} = \infty, \;
\kappa_{\rm tric} = 0.10, \;
\lambda_{\rm tric} = 0 \,\,.
\end{eqnarray}
Since this result may be an artefact of the approximations involved,
we look for the tricritical point in the linear
approximation of $\beta_{e^2}$ -- this corresponds to eq.~(\ref{e2flow}) --
where the effects of the mass of the fluctuations
are ignored and $\ell_{\rm eff}=\ell_{gc}$.
This approximation strengthens the running of the gauge
coupling, thus resulting in a lower value for
$e^2_\star$. In this linear approximation, the tricritical
fixed point now appears at
\begin{eqnarray}
&\varphi^4:&  \; \; \; \; \; e^2_{\rm tric} = 70.4, \; \; \;
\kappa_{\rm tric} = 0.434,   \; \; \;
\lambda_{\rm tric} =  0.027, \\[1ex]
&\varphi^8:&  \; \; \; \; \; e^2_{\rm tric} = 70.4, \; \; \;
\kappa_{\rm tric} =   0.157, \; \; \;
\lambda_{\rm tric} =  1.97,
\end{eqnarray}
while the second-order fixed point gets shifted
from the values (\ref{fpphi4}-\ref{fpphi8}) to
\begin{eqnarray}
&\varphi^4:&  \; \; \; \; \; e^2_{\star} = 70.4, \; \; \;
\kappa_{\star} =  0.057,   \; \; \; \lambda_{\star} =  20.2, \\[1ex]
&\varphi^8:&  \; \; \; \; \; e^2_{\star} =  70.4, \; \; \;
\kappa_{\star} =  0.124,  \; \; \; \lambda_{\star} =  3.73~.
\end{eqnarray}
In any case, the tricritical point characterizes the boundary between
first- and second-order behaviour and we cannot be sure that a polynomial
expansion of $u_k$ gives a quantitatively reliable result.
This is underlined by the small value of $\lambda_{\rm tric}$ as compared
to $\lambda_\star$, in particular in the $\varphi^4$-approximation.
The strong decrease of $\lambda_\star$
between the $\varphi^4$-
and the $\varphi^8$-approximation
may also be viewed as a source of
worry about the polynomial expansion
at the second-order fixed point if the linear approximation
with $e^2_* = 70.4$ is used.
The phase diagram for the $\varphi^4$-approximation
with $\beta_{e^2}$ given by
the linear approximation (\ref{e2flow}) is shown in Figs.~2 and 3.

Further evidence for the existence of a second-order fixed point
comes from the investigation of an abelian Higgs model with $N$
complex scalar fields and an extrapolation to $N=1$.
One expects that in the presence of a large number of scalar fields $N$
the role of the gauge field is diminished,  and
the description of the scaling behaviour
of the system within our approximation is highly precise.
For this reason, it is of particular interest to cross-check
our results by investigating the scaling solutions
of a system with $N$ scalar fields and to look for the stability of the
behaviour as $N$ goes to one. In this way, the large $N$ limit of
the theory provides an attractive way
to check the consistency of our findings.

In~\cite{largeN} the $N$-dependent form of the evolution equations
for the scalar and gauge couplings, as well
as for the minimum of the potential, is derived
using  the classical action for the gauge field and an
$O(2N)$ symmetric scalar sector. This symmetry is reduced to
$SU(N) \times U(1)$ due to the coupling to
the photon. The equations read in the $\varphi^4$-approximation
\begin{eqnarray}
\frac{de^2}{dt}\!\!&=&\!\!\beta_{e^{2}} = -e^2
+\frac{4}{3}v_3 e^4 \left[\ell_g(2 \lambda \kappa, 2e^2\kappa)
+ \ell_c(2 \lambda \kappa) +(N-1)\ell_{gc} \right]
\label{betae2} \\
\nonumber\\
\frac{d \kappa}{d t}\!\!&=&\!\!\beta_{\kappa}~\!=
\left( 1+ \eta_\varphi \right) \kappa +
      8 \frac{e^2}{\lambda}v_3
      \ell_1(2 e^2 \kappa)+6 v_3 \ell_1(2 \lambda \kappa)
      + 2 (2N-1) v_3 \ell_1  \label{betaka} \\
\nonumber\\
\frac{d \lambda}{d t}\!\!&=&\!\!\beta_{\lambda}~\!=
\!-\left( 1\! -\! 2 \eta_\varphi \right) \lambda
                    + 16 e^4 v_3 \ell_2 (2 e^2 \kappa)+18 \lambda^2
      v_3 \ell_2 (2 \lambda \kappa) + 2(2N\!-\!1)\lambda^2 v_3 \ell_2
\label{betala}
\end{eqnarray}
where $\eta_\varphi$ is as in (\ref{scandim}).
We see that the only modification with respect to
$\beta_{\kappa}$ and $\beta_{\lambda}$
in eqs.~(\ref{kappaflow}-\ref{lambdaflow})
arises from diagrams that involve the $(2N-1)$ massless
modes inside the loop, while for $\beta_{e^2}$ the $N$-dependence
arises from the graphs which involve the $(N-1)$ massless complex scalar
fields.

These equations indicate that, for a large number of scalar
fields $N$, we may obtain the fixed points of the theory
by considering only the leading $N$-dependent
contributions in the $\beta$-functions.
In this case, the evolution equation
for the gauge coupling decouples
from the scalar sector and
has a non-trivial infrared fixed point solution,
given by
\begin{equation}
e^2_\star=\frac{3}{4 v_3 \ell_{gc}}\frac{1}{N}=\frac{70.4}{N}.
\label{e2f}
\end{equation}
Note that in leading order in $1/N$-expansion we recover the value for
$e^2_\star$ in the linear approximation for $N=1$.
The second-order fixed points for the scalar sector may be obtained by
expressing the large $N$ behaviour of the couplings as
$\kappa \sim N, \lambda \sim N^{-1}$.
The influence of the gauge coupling and of
the massive scalar fluctuations in the evolution equations is of
subleading order in $1/N$ and the fixed point values are entirely
determined through the fluctuations of the massless modes to be
\begin{eqnarray}
\kappa_\star = 4 v_3 \ell_1\,N \label{ka1f}, \; \;  \; \; \; \;
\lambda_\star =  \frac{1}{4 v_3 \ell_2}\, \frac{1}{N}.\label{la1f}
\end{eqnarray}
The tricritical fixed points arise due to the presence of the gauge field,
and the relevant solutions are more complicated. In this case one
finds~\cite{largeN}
$\kappa\sim N$, $\lambda\sim N^{-2}$ and
the dependence of $\lambda$ on the number
of scalar fields is much stronger for the tricritical point than the
second-order one.

The $N$-dependence of the fixed points
may be discussed in terms of a simple differential
equation which uses a linearization
of the flow equation for couplings near the fixed point
values
\begin{equation}
\frac{\partial \vec{\lambda}_{\star}}{\partial N}(N)
= - A^{-1}( \vec{\lambda}_{\star}(N),N)
\frac{\partial \vec{\beta}}
{\partial N}
(\vec{\lambda}_{\star}(N),N)
\label{flo}
\end{equation}
where
\begin{eqnarray}
\vec{\lambda} =
\left (
\begin{array}{c}
\kappa \\
\lambda \\
e^2
\end{array}
\right), \; \; \; \; \;
\vec{\beta}(\vec{\lambda}(N),N) =
\left (
\begin{array}{c}
\partial_t \kappa \\
\partial_t \lambda \\
\partial_t e^2
\end{array}
\right), \; \; \; \; \;
A(\vec{\lambda},N)=\frac{\partial \vec{\beta}}{\partial \vec{\lambda}}
(\vec{\lambda},N).
\end{eqnarray}
This equation has a solution
as long as the stability matrix $A$ of the system
(which involves the
derivatives of the $\beta$-functions
at the fixed point) is regular.
It was found that
all eigenvalues of the stability
matrix depend only moderately on $N$,
and remain far away from zero
down to $N =1$. This indicates that
a second-order fixed point and the associated
parameter region for a second-order phase transition exists for
all values $N \geq 1$.
Solving the differential equation (\ref{flo}) it was indeed observed that
the system (\ref{betae2}-\ref{betala}) always
has non-trivial fixed points. These results are in perfect
agreement with the numerical solution of the
renormalization group equations that describe the
evolution of the system for arbitrary $N$.
Therefore, the investigation of the fixed point structure of
the abelian Higgs model in three dimensions,
using both a semi-analytical
and a numerical approach, points to the
existence of a region with a second-order phase transition
down to $N=1$.

Concerning the $N$-dependence of
the critical exponents, it was found that
$\eta_\star$  is given by
\begin{equation}
\eta_\star=\left[ \frac{4}{3} \frac{\ell_1}{(\ell_2)^2}
\, m_{2,2}(\tilde{m}^2_\star,0) -
\frac{8}{\ell_{gc}}\, \ell_{1,1}(\tilde{m}^2_\star,
\tilde{M}^2_\star) \right] \frac{1}{N}
\label{etaf1}
\end{equation}
where $\tilde{m}_\star$ and $\tilde{M}_\star$
stand for the dimensionless scalar and gauge field mass
at the fixed point, respectively
($ \tilde{m}^2_\star = 2 \lambda_\star \kappa_\star =
2\ell_1 / \ell_2$, $ \tilde{M}^2_\star = 2
e^2_\star \kappa_\star =  6 \ell_1 /\ell_{gc}
$).
The first term in eq.~(\ref{etaf1})
accounts for the scalar fluctuations only, and coincides with
the result for the pure scalar theory in the truncation used here
\cite{critexp}.
The second term arises due to the
gauge field fluctuations and gives the dominant
contribution to $\eta_\star$,
which is therefore negative: $\eta_{\star}=
-\frac{0.31}{N}$. The critical index $\nu$ behaves like
$1-{\cal{O}}(N^{-1})$ for large $N$. We have obtained this
number numerically to be $\nu_{\star} =
1-\frac{1.38}{N}$.

These results may be compared with those found by alternative
methods. For example results obtained using the $\epsilon$-expansion
give a leading $1/N$ coefficient $-9\epsilon$ for
$\eta$
whereas for the subleading coefficient of $\nu$,
$N(\nu-1)$, values range from
$-2\epsilon$~\cite{nematic} to $-48\epsilon$~\cite{HLM}.
These numbers indicate that both exponents run
in the unphysical region already for
large $N$. The large
$N$ results which have been obtained directly in three dimensions
predict
a value $-2.21$ and $-4.86$ for the $1/N$ coefficients
of $\eta$ and $\nu$ respectively~\cite{{HLM},{nematic}}.
These results, although exact in the limit
$N \rightarrow \infty$, are less accurate as $N$ decreases
and for $N=1$, where they indicate a first-order transition, the
deviation is maximal. Nevertheless, we can compare these predictions
with our expectations in order to understand the
limitations of our approximations. Since for $\eta_\star$ we have
analytic expressions which are directly associated with diagrams,
we focus on this exponent. As we see, there is a disagreement between
our findings and the previous results for $N \rightarrow \infty$.
To understand the source of the deviation, we first compare the
 contributions of the scalar fluctuations only.
This part has been calculated in \cite{critexp}
and gives a term $0.110/N$ to be compared with
$0.135/N$, obtained using large N
techniques in three dimensions. We found that the larger
part of this deviation is due to the non-inclusion of
$Y_{k}$ (cf.~eq.~(\ref{sechzehn})) in our
truncations. Once this is incorporated,
we instead get a term $0.129/N$.
The remaining deviation for $N \rightarrow \infty$
most probably arises due to neglecting the
momentum dependence.
For the gauge fluctuations, the situation is more
complicated: here the effects of dropping the
momentum dependence are larger,
as we can see by looking
at the
wave function renormalization.
This scales as
$Z \sim (k^2+q^2)^{-\eta/2}$, thus for
large $\eta$ the momentum dependence of
the results is not negligible.
For the gauge field, where the anomalous
dimension is larger, the effect is more significant
and the corrections to $Z_{\scriptscriptstyle F}$ are presumably
bigger than those to $Z_{\varphi}$.
However, when calculating the anomalous
dimension $\eta_{\varphi}$,
$\frac{\partial Z_{\scriptscriptstyle F}}{\partial q^2}$
enters the calculation leading to an error
in the $1/N$ contribution to the anomalous dimension.

We see therefore that for a better understanding of the phase transition
we should also include the momentum dependence
of $Z_{\scriptscriptstyle F}$.
Nevertheless, while our present truncation has limitations,
it also has certain advantages as compared to the standard
large $N$ results, namely for
medium and small values of $N$. In our approach the threshold effects
due to the decoupling of massive fluctuations, which become very important
as $N$ decreases, are included. The relevance of such effects may be
seen from the deviation of the critical indices that we calculate
explicitly as compared to the ones that we would expect from the
large $N$ limit of our method.
In the latter case the effect of the
masses in the calculation of $\kappa_\star$,
$\lambda_\star$ and $e^2_\star$ is neglected.
For both
$\eta$ and $\nu$ we see that these threshold effects result in an
important increase of the indices, which become compatible with a
second-order phase transition.
For example, for $N=1$ our naive large $N$ estimates
indicate a negative $\nu$, while the numerics, taking into account
threshold effects, give an important correction driving the value to
$0.532$ and $0.583$ in the $\varphi^4$- and $\varphi^8$-approximations
respectively. These values are well within the range of a second-order
transition, and that gives information on the implications
of threshold effects.
Actually, the critical indices
are in the same range as those measured in the nematic-to-smectic-A
transition (see section 8).
We emphasize that the error in the $1/N$-coefficients for large $N$
concerns already subleading terms for the behaviour at the fixed
point. There is no reason to believe that the error due to
our truncations scale $\sim N^{-1}$ as $N$ becomes small.
Of course definite conclusions should be drawn once the
momentum dependence are
included. However, this first attempt to apply the method of
average action in this problem seems to offer substantial
improvement in certain directions, like the inclusion of threshold effects
that we have discussed,
and encourages a more elaborate analysis which can give an improved
picture of the transition.
Note that the error in $\eta$ does not affect
our numerical searches for fixed points, since the deviations
are subdominant as compared to the
rest of the contributions.

At this stage we can also look at the
$N$-dependence of the tricritical fixed point,
which has been a delicate
point in our analysis. It turned out that for $N=1$ there is a
region in parameter space where
$\eta_{\scriptscriptstyle F}$ (eq.~(\ref{ggandim})) turns
negative. For negative $\eta_{\scriptscriptstyle F}$,
$\beta_{e^2}$ is negative for all $e^2$
and this coupling runs to infinity. In Fig.~9, the dashed line
indicates the border between negative (above) and positive (below)
$\eta_{\scriptscriptstyle F}$ for $N=1$
and $\kappa$ given by eq.~(\ref{fpphi4}). This
region is compared with the tricritical fixed points (left branch)
and the second-order fixed points (right branch) given in the
$\varphi^4$- (stars) and the $\varphi^8$- (diamonds) approximation
for various values of $N$. One clearly sees how for $N\to1$
the tricritical fixed points  approach the region of negative
$\eta_{\scriptscriptstyle F}$ for $N=1$. In contrast, the second-order fixed
points remain always within the reliable region.

In summary, we have given a numerical analysis of the phase diagram
of superconductors based on non-perturbative flow equations. Their
derivation involves approximations whose validity still needs a careful
checking -- see section 7. Within our truncations we establish the existence
of a parameter region where the phase transition is second-order.
The $1/N$-expansion gives further evidence for
the existence of non-trivial fixed points -- and therefore the existence of
a parameter region where the phase transition is second-order -- for any $N$,
in contrast to previous results where a large number
of complex fields was required for this purpose.
We believe that the difference arises because
in our approach the threshold effects due
to the decoupling of massive fluctuations are
taken into account.

\end{section}


\begin{section}{\names Field Dependent Gauge Coupling}

The central equation (\ref{flowpot}) of our approach
describes the scale dependence of the coarse grained
free energy $u_k(\tilde{\rho})$ in dimensionless units.
This is a partial differential equation
for the function $u$ depending on two variables $t$ and $\tilde{\rho}$,
which can be solved numerically.
The problem here is the unknown function
$z_{\scriptscriptstyle F}(\tilde{\rho})$.
If one puts $z_{\scriptscriptstyle F}(\tilde{\rho}) = 1 $
and uses eq.~(\ref{e2flow}) for the running of $e^2$ no scaling
solution is found \cite{private}.
Similar results have been obtained in~\cite{largeN} through an
algorithm introduced
in~\cite{4D}. Both methods therefore do not reveal a second-order
fixed point for $N=1$ whereas a scaling solution is indeed
observed for $N~\geq~4$.
In contrast to the results of the last section these findings seem to suggest
that for $N=1$ the phase transition is always first-order. They also may cast
doubts about the validity of the local approximation employed there.

The main reason for the absence of a scaling solution is that the very large
value of $e^2 \approx e^2_\star$ always drives the mass
term at the origin to positive values in the critical region,
as can be seen from the flow equation for $\tilde{m}^2 = u'_k(0)$
\begin{eqnarray}
\partial_t \tilde{m}^2 =
-(2-\eta_{\varphi})\tilde{m}^2 -
8v_3 u''_k(0) \ell_1(\tilde{m}^2)
- 8 v_3 e^2 \tilde\ell_1(0,z_{\scriptscriptstyle F}(0))
- 4v_3 z_{\scriptscriptstyle F}'(0)
\tilde\ell^5_1(0,z_{\scriptscriptstyle F}(0))~~~~
\end{eqnarray}
where the functions $\tilde\ell^d_n$ are given in the appendix.
For $z_{\scriptscriptstyle F}(0) =1$,
$z_{\scriptscriptstyle F}'(0) =0$ and $u''_k(0)>0$
the large negative term $\sim e^2$
always dominates. This is incompatible with a negative constant
$\tilde{m}^2$ which would be required for the
scaling solution at a second-order phase transition.
Note that $\tilde{m}^2$ must be larger than
$-1$ as all arguments of threshold functions.
A scaling solution
requires
\begin{eqnarray}
\tilde m^2_\star = -\frac{8 v_3}{2+\eta_{\varphi \star}}
\left ( u''_\star(0) \ell_1(\tilde{m}_\star^2) +
e^2_\star \tilde\ell_{1}(0,z_{{\scriptscriptstyle F} \star}(0)) +
\frac{z'_{{\scriptscriptstyle F} \star}(0)}{2}
\tilde\ell^5_1(0,z_{{\scriptscriptstyle F} \star}(0))
\right ) > -1
\label{mass}
\end{eqnarray}
which for $\tilde z_{\scriptscriptstyle F}(\rho)=1$ is
possible only for large negative values
of $u''_\star(0)$. Such a negative quartic coupling for $\tilde\rho=0$
rapidly destabilizes the system and destroys a possible scaling solution.
Another way to fulfill the condition~(\ref{mass}) arises for
$z_{{\scriptscriptstyle F} \star}(0)$
sufficiently large. We will argue in the following
that generically $z_{\scriptscriptstyle F}(0)$ is indeed a large quantity.

Let us define an effective $\rho$-dependent
gauge coupling by
\begin{eqnarray}
e^2(\tilde{\rho}) =
\frac{\bar{e}^2}{Z_{{\scriptscriptstyle F},k}(\tilde\rho) k} =
\frac{e^2}{z_{\scriptscriptstyle F}(\tilde{\rho})}~.
\end{eqnarray}
This is the quantity which describes the effective
coupling between the scalar and the gauge field
for arbitrary background field $\rho$, with
$e^2 = e^2(\kappa)$.
The running of the function
$e^2(\tilde{\rho})$ is given by
\begin{equation}
\partial_t e^2(\tilde{\rho}) = -
e^2(\tilde{\rho}) + \eta_{\scriptscriptstyle F}(\tilde{\rho})
e^2(\tilde{\rho})
\end{equation}
where
\begin{equation}
\eta_{\scriptscriptstyle F}(\tilde{\rho}) = - \partial_t
\ln Z_{{\scriptscriptstyle F},k}(\tilde\rho).
\end{equation}
A scaling solution can only occur for
\begin{equation}
\eta_{{\scriptscriptstyle F} \star} (\tilde{\rho}) = 1
\end{equation}
A full calculation of
$\eta_{\scriptscriptstyle F} (\tilde{\rho})$,
which is equivalent to the evolution
equation for $Z_{\scriptscriptstyle F} (\tilde{\rho})$,
is not yet undertaken here.
For an order of magnitude estimate we
approximate $\eta_{\scriptscriptstyle F} (\tilde{\rho})$ by neglecting
on the right-hand side the $\tilde{\rho}$ derivatives of
$Z_{\scriptscriptstyle F} (\tilde{\rho})$.
Then $\eta_{\scriptscriptstyle F} (\tilde{\rho})$ can
simply be inferred from~\cite{NPB427} by replacing in $\ell_g$ the
arguments by appropriate $\tilde{\rho}$ dependent mass terms. One finds
\begin{eqnarray}
\eta_{\scriptscriptstyle F}(\tilde{\rho})&=&
\frac{32 \ell_g}{15 m_4^5}~
v_3 e^2(\tilde{\rho})~m_{2,2}^5
( u'(\tilde{\rho})
+ 2 \tilde{\rho}u''(\tilde{\rho}),
u'(\tilde{\rho}) ) \\
&+&\frac{64}{15}
v_3 e^4(\tilde{\rho})\tilde{\rho}~\big[4 m_{2,2}
( u'(\tilde{\rho})
+ 2 \tilde{\rho}u''(\tilde{\rho}),
 2 e^2 (\tilde{\rho}) \tilde{\rho})
-n_{2,1}^1
( u'(\tilde{\rho})
+ 2 \tilde{\rho}u''(\tilde{\rho}),
 2 e^2 (\tilde{\rho}) \tilde{\rho})\big]\nonumber
\end{eqnarray}
The various threshold functions can be found in the appendix and we
have again suppressed the separate dependence of some of these functions
on $z^{-1}_{\scriptscriptstyle F}(\tilde\rho)$ (cf. the discussion
of $\tilde \ell_0$). The $\tilde{\rho}$ dependence of
the mass terms which appear in the arguments
of the threshold functions actually  plays an important
role. This can be seen by looking at the scaling
solution $\eta_{{\scriptscriptstyle F} \star} = 1$:
the function $e^2_\star(\tilde\rho)$ is determined by this
condition in complete analogy to
$e^2_\star = e^2_\star({\kappa_\star})$,
and the $\tilde{\rho}$-dependence of this function arises
only by the difference of the scalar mass terms
$u'(\tilde{\rho}) + 2 \tilde{\rho}u''(\tilde{\rho})$
and $ u'(\tilde{\rho})$ as compared to the
minimum values $2\lambda \kappa$ and $0$.
Especially, we expect a negative value $\tilde{m}^2_\star =
u_\star'(0)$ for the mass term at the origin for a second-order
fixed point. This
negative value considerably increases the values of the
threshold functions and therefore leads to a small value
of the ratio $e^2(0)/e^2$. Since this ratio multiplies
the gauge boson contributions to the flow of the average potential around
the origin the situation described at the beginning of this section
presumably gets strongly modified once the
$\tilde{\rho}$-dependence
of $e^2$ is taken into account.
We conclude that a numerical solution of the
full differential equation for
$\partial_t u_k(\tilde{\rho})$ will only make
sense if it is coupled to
a similar equation for
$\partial_t e^2(\tilde{\rho})$.
Without an inclusion of the $\rho$-dependence
of $Z_{\scriptscriptstyle F}$ the truncations
discussed in section 4 give probably
a more realistic picture than the full equation
(\ref{flowpot}), since only a relatively small range of
$\tilde{\rho}$ is relevant in this case.

\end{section}


\section{\names Discussion and Conclusions}
In this paper we have presented a phase diagram
(Figs.~2 and 3) for the phase transition to
superconductivity. On the critical surface we
observe two distinct regions, one leading to a
first-order phase transition and the other to the scaling
behaviour of a second-order transition.
A relatively
large
discontinuity can only be observed for good
type-I superconductors with small values of
$\bar{\lambda}(\Lambda)/\bar{e}^2(\Lambda)$.
On the other hand, second-order behaviour is
expected for good type-II superconductors, where
$\bar{\lambda}(\Lambda)$ is much larger than
$\bar{e}^2(\Lambda)$.
The quantitative details of the phase transition are
not equally well understood for all parts of the
phase diagram.
This is related to the question of the validity of
the truncations which were necessary in order to
extract our system of non-perturbative flow equations
from the exact evolution equation for the average
action.
Nevertheless, we emphasize that the flow equations
used in this paper contain information which goes
far beyond all presently available approaches. We
can work directly in arbitrary dimension $d$ and
cope with the infrared problems related to massless
excitations.
The loop expansion or the $\epsilon$-expansion around
$d=4$ can be viewed as
rather
rough approximations to our equation, which, in particular,
neglect the important threshold effects related to the
masses of the excitations. This explains the failure
of these methods to describe the critical scaling
behaviour for large values of
$\bar{\lambda}(\Lambda)$.
Neglecting scalar fluctuations,
the one-loop result is recovered from our
equation if the running
of $e_{\scriptscriptstyle R}^2$ is neglected.
The one-loop result including scalar fluctuations
obtains by approximating on the right-hand side
of eq.~(\ref{zweiundzwanzig})
$U_k'(\rho) = -\bar{\mu}^2$ and
$U_k''(\rho) = \bar{\lambda}(\Lambda)$.
On the other hand, our flow equations can
easily be formulated for arbitrary dimension $d$.
For $\epsilon\rightarrow 0$ they reproduce the
results of the $\epsilon$-expansion, since in this
limit the mass terms $2 \lambda_\star \kappa_\star$
and $2 e^2_\star \kappa_\star$ appearing in
the threshold functions are small quantities
$\sim \epsilon$. In lowest order in the
$\epsilon$-expansion, equation (\ref{betaratio}) contains
all relevant information once the constants like
$\ell_2$ and $\ell_{\rm eff}$ are replaced by
appropriate quantities for $d=4$.
No fixed point is present and the transition
would be first-order for all values of
$\bar{\lambda}$ and $\bar{e}^2$.
Schwinger-Dyson or gap equations can also be
interpreted as solutions of the flow equations.
In particular, the gap equation for the mass term
obtains by replacing $\bar{\mu}^2$ by some
unknown parameter $\mu^2$ which is then equated
self-consistently with $-U'(0)$ for $k=0$.
Similarly, the self-consistent screening
approximation used in reference~
\cite{rad}
may be viewed as a generalization of the gap equation
for a momentum dependent renormalized propagator
and quartic scalar self-coupling.
It follows from the exact evolution equation for the
two- and four-point function (not displayed here)
in the limit where the higher $n$-point functions
are neglected and the $k$-dependent couplings
are replaced
on the right-hand side of the flow equation
by $k$-independent couplings evaluated for $k=0$
in a self-consistent way.
This method of equating short distance couplings
with renormalized (long distance) couplings is
not a very accurate approximation to the solution
of flow equations in situations where the running
of couplings has an important effect.
Nevertheless, this method has led to a qualitatively
similar picture to ours for the existence of a
parameter range with second-order behaviour~\cite{rad}.
In particular a value of $-0.38$ is found for $\eta_\varphi$,
which is smaller than our values shown in Table 1.

Our flow equations are known~\cite{critexp}
to give a very accurate description of the
$O(N)$-symmetric Heisenberg model. Also the
running of $e^2(k)$ in the region of
small $e^2$ is quantitatively reliable. The
truncated terms only lead to corrections in the
$\beta$-function for $e^2$ that involve higher
powers of $e^2$, like $e^6$ etc. The part of
the phase diagram with small $e^2$ and
arbitrary $\lambda$ is therefore highly reliable.
Another part which is
quantitatively
well understood concerns the region of small
$\lambda/e^2$ (see section~5). The main uncertainty
for the flow of the couplings in this region of
parameter space concerns the quantitative determination
of the fixed point value for the dimensionless
gauge coupling $e_\star^2$.
In contrast to the case with a larger number of
scalar fields we do not believe that we can
determine this value accurately for $N=1$
within our truncation.
A quantitative control requires at least the
inclusion of the $\rho$-dependence of the
gauge coupling as mentioned in section 7,
and probably also of the momentum dependence of $Z_{\scriptscriptstyle F}$.
Nevertheless, for small $\lambda/e^2$ the
relative change in $\lambda$ is much more
important than the relative change in $e^2$,
independently of the precise value of $e_\star^2$.
The results concerning the first-order transition
for small $\lambda/e^2$ are therefore very
reliable.

Next we turn to the region in parameter space in the
vicinity of the second-order fixed point. We have
computed this fixed point by a polynomial expansion
of the free energy around its minimum.
We have used different truncations, and the apparent
convergence is satisfactory for the $\varphi^8$- and
higher approximations. Also the positive
eigenvalues of the stability matrix for the flow
of small deviations from the fixed point are of order
one~\cite{largeN},
indicating a rather high stability of the system with
respect to perturbations. The mass terms of scalar and
gauge boson fluctuations at the fixed point are large
compared to $k^2$ and the $\varphi^4$-coupling
$\lambda_\star$ is substantial. All this seems to
suggest that an expansion around the minimum of the
free energy gives a meaningful result.
For a quantitative determination of the critical
behaviour, the fixed point $e_\star^2$ again appears
as the weakest part. Fortunately, critical exponents
and other universal quantities do not depend strongly
on $e_\star^2$ if $e_\star^2 \kappa_\star$ is large
enough. We give in table 1 the values of the exponents
$\nu$ and $\eta$ for three different assumptions for
$e_\star^2$: a) the linear approximation which is
equivalent to our leading $1/N$ result in $\beta_{e^2}$, b) the
solution of the flow equations
(\ref{flowe2}) and (\ref{ggandim}), and c) the limit
$e_\star^2\rightarrow \infty$ as obtained from the
fixed point of the corresponding non-local
scalar model.
\begin{table}
\centering
\begin{tabular}
{c|cc|cc|cc}
\hline \hline
 & \multicolumn{2}{c|}{(a)} &  \multicolumn{2}{c|}{(b)} &
 \multicolumn{2}{c}{(c)}\\[.5ex]
 & $\varphi^4$ & $\varphi^8$ &$\varphi^4$ & $\varphi^8$ &
$\varphi^4$ & $\varphi^8$ \\[1ex]
\hline
$\eta$ & --0.13 & --0.20 & --0.13 & --0.17 & --0.13 & --0.15 \\[.5ex]
$\nu$ & ~~0.50 & ~~0.47 & ~~0.53 & ~~0.58 & ~~0.59 & ~~0.62 \\[.5ex]
\hline \hline
\end{tabular}
\begin{center}
{\bf Table 1}
\end{center}
\end{table}

We indicate results from the $\varphi^4$- and
$\varphi^8$-approximation separately. We see
that the critical exponents are rather robust with
respect to these different approximations and we find
$\eta$ between $-0.20$ and $-0.13$ and $\nu$ between
$0.47$ and $0.62$. From the scaling laws we infer
$\alpha$ between
$0.14$
and
$0.59$,
$\beta$ between
$0.19$
and
$0.27$,
and $\gamma$ between
$1.00$
and
$1.36$.
It has been proposed~\cite{gennes},
that the critical behaviour for the
nematic-to-smectic-A transition
in liquid crystals is described by the same
universality class as the normal-to-superconductor
transition. In these systems a critical behaviour
consistent with a second-order phase transition
has been observed and the critical exponents are
measured.
Comparison with the experimental values is not straightforward
due to the fact that the liquid crystals exhibit
anisotropic scaling.
A recent compilation of Garland and Nounesis
\cite{newdata}
gives values
$\nu_\parallel = 0.54 - 0.83$,
$\nu_\perp = 0.37 - 0.68$,
$\gamma = 1.07 - 1.53$, and
$\alpha = -0.03 - 0.50$. It is encouraging that these
ranges roughly coincide with our estimates.
However, one should keep in mind
that it is not obvious that all observed
systems show the critical behaviour related to
the second-order fixed point or instead the
one related to some crossover phenomenon
(see below).
This holds in particular for the substances with small
$\nu$ and $\gamma$ and large $\alpha$.

The less well understood region in the phase diagram
concerns first the behaviour for $e^2\rightarrow \infty$.
For large $e^2$, the validity of the truncations
leading to eqs.~(\ref{flowe2}) and (\ref{ggandim}) is doubtful
and, in particular, the $\rho$-dependence of the
effective gauge coupling becomes an important
effect. A more detailed study of the non-local
scalar theory which is obtained in for
$e^2\rightarrow \infty$, with truncations better
adapted for this case than the ones used in this paper,
may provide a reliable answer on the $\beta$-function
for large $e^2$.
Related to this problem is our poor understanding
of the tricritical fixed point. We see this fixed point
in certain approximations for the running of $e^2$
(eq.~(\ref{e2flow})), but not in others (eqs.~(\ref{flowe2}) and
(\ref{ggandim})). No quantitative conclusions can therefore
be drawn concerning the crossover behaviour from the
tricritical fixed point.
Finally, in direct relation with the uncertainties at
the tricritical fixed point is our inability to give
a precise estimate of the critical ratio
$\left( \lambda(\Lambda)/e^2(\Lambda) \right)_c$, for which the
phase transition changes from first- to second-order.
We have given an order of magnitude estimate (\ref{approxratio})
and approximate lower (Fig.~4) and upper (eq.~(\ref{g76}))
bounds. Again, we hope that a correct treatment of
$Z_{{\scriptscriptstyle F},k}(\rho)$ will improve the situation considerably.

There also remain questions about the overall picture
sketched in the phase diagrams (Figs.~2 and 3).
This concerns the absence of a scaling solution of the
system of differential equations (\ref{flowpot}) and (\ref{e2flow})
with $z_{\scriptscriptstyle F}(\tilde{\rho})=1$, $\eta_\varphi=0$.
As discussed in section 7, the true answer about the
existence of a scaling solution requires the understanding
of the function $e^2(\tilde{\rho})$.
A too large value of $e^2(0)$ inevitably produces a minimum
of the free energy at the origin in the vicinity of the
critical temperature, and therefore always turns the
phase transition first-order. The crucial question for
a confirmation of our picture of first- and second-order
regimes concerns the true value of $e^2(0)$. Only for
a fixed point value $e_\star^2(0)$ smaller than a critical
value is a second-order phase transition possible.
Otherwise the transition is first-order for all
parameter values $e^2(\Lambda)$ and $\lambda(\Lambda)$.

Several predictions about possible observations of critical
behaviour are, however, independent of the unresolved
details of the phase transition.
They can be tested by experiment or lattice simulations.
Once $\lambda(\Lambda)/e^2(\Lambda)$ becomes
sufficiently large, one should start seeing universal scaling
behaviour in the vicinity of the critical temperature.
The effective critical exponents which may be measured close,
but not too close, to the critical temperature will first
reflect the crossover from the gaussian or the Wilson-Fisher
fixed point, depending on the value of
$\lambda(\Lambda)/e^2(\Lambda)$.
For the borderline between type-I and type-II superconductors
where $\lambda(\Lambda)\simeq e^2(\Lambda)$, one should
find mean field exponents for a first observation of the
approximate scaling behaviour. For
$\lambda(\Lambda)\simeq 10 e^2(\Lambda)$, the first observation
of scaling behaviour in real superconductors should see the
critical exponents of the (two-component) Heisenberg model. Only
for still larger $\lambda(\Lambda)/e^2(\Lambda)$
and very close to the critical temperature the exponents of the
new second-order fixed point may be detected.
The situation is different for liquid crystals and lattice
simulations, where larger values of $e^2(\Lambda)$
may be realized. If a region in parameter space with a second-order
phase transition exists, we believe that our estimates for the ranges
of critical exponents are rather reliable. Lattice simulations
should be able to test our predictions.
The interpretation of observed critical exponents in liquid
crystals has to be done with care. It is possible
that for certain materials the effective critical exponents
describe the crossover from the Wilson-Fisher fixed point
or reflect even more complicated anisotropic fixed points.
This may be responsible for the fact that the data for the
critical indices for different materials does not always
coincide very well.

\bigskip
\bigskip

\noindent{\names Acknowledgments}

\bigskip

We would like to thank N. Tetradis for numerous interesting discussions
about this work.

\vfil
\eject




\begin{appendix}
\noindent{\names Appendix: Threshold Functions and Constants}

\bigskip

\noindent In this appendix we give the threshold functions and
constants used in this paper. We will here give the
expressions in arbitrary dimensions, while in the
main text $d=3$ is understood if not explicitly
denoted otherwise.

We have:
\begin{eqnarray}
l^{d \geq 3}_g  \!&=&\!
     - \frac{d-2}{4} k^{4-d} \int_0^{\infty}
     dx~x^{\frac{d}{2}-2} \frac{d}{dx} \left[
     \frac{1}{P(x)} \tilde{\partial_t} P(x)
     \right] \nonumber\\
\nonumber\\
l^d_{0}(\omega) \!&=&\!
      \frac{1}{2} k^{- d} \int_0^{\infty} dx~
     x^{\frac{d}{2}-1}
     \tilde{\partial}_t \ln\left(
    P(x)+ \omega k^2 \right)  \nonumber \\
\nonumber\\
l^d_{n \geq 1}(\omega) \!&=&\!
     - \frac{1}{2} k^{2 n - d} \int_0^{\infty} dx~
     x^{\frac{d}{2}-1}
     \tilde{\partial}_t \left(
    P(x)+ \omega k^2 \right) ^{-n} \nonumber \\
\nonumber\\
l^d_{n,m}(\omega_1, \omega_2) \!&=&\!
     - \frac{1}{2} k^{2 (n+m) - d} \int_0^{\infty} dx~
     x^{\frac{d}{2}-1} \times \nonumber \\
     & &\! \tilde{\partial}_t
     \left[ \left( P(x) + \omega_1 k^2
          \right)^{-n}
     \left( P(x) + \omega_2 k^2
          \right)^{-m} \right] \nonumber\\
\nonumber\\
m^d_{n,m}(\omega_1, \omega_2) \!&=&\!
     - \frac{1}{2} k^{2(n+m-1)-d} \int_0^{\infty} dx~
     x^{\frac{d}{2}} \times \nonumber \\
     & &\!  \tilde{\partial}_t \left[
     \left( \frac{dP}{dx} \right)^2
     \left( P(x) + \omega_1 k^2
          \right)^{-n}
     \left( P(x) + \omega_2 k^2
          \right)^{-m} \right] \nonumber\\
\nonumber\\
n^d_{n,m}(\omega_1, \omega_2) \!&=&\!
     - \frac{1}{2} k^{2(n+m-1)-d} \int_0^{\infty} dx~
     x^{\frac{d}{2}} \times \nonumber \\
     & &\!  \tilde{\partial}_t \left[
     \frac{dP}{dx}
     \left( P(x) + \omega_1 k^2
          \right)^{-n}
     \left( P(x) + \omega_2 k^2
          \right)^{-m} \right] \nonumber\\
\nonumber\\
l^{d \geq 3}_c(\omega)  \!&=&\!
     \frac{d-2}{4} k^{4-d} \int_0^{\infty}
     dx~x^{\frac{d}{2}-2} \frac{d}{dx} \frac{d}{dt}
     \ln \left( 1 + \frac{P(x)-x}{k^2(1+\omega)}
     \right) \nonumber \\
\nonumber\\
l^d_{gc}\!&=&\!l^d_g +l^d_c(0) .
\label{A.1}
\end{eqnarray}
Here
$\tilde{\partial}_t$
stands for
$\frac{\partial}{\partial t}$
acting only on $R_k$ contained implicitly in $P(x)$  (see sect.~3).
The threshold functions appearing in the main text are easily related
with the ones in the above list. Explicitly:
$\ell_g=l^3_g$, $\ell_n=l^3_n$, $\ell_{1,1}=l^3_{1,1}$,
$m_{2,2}=m^3_{2,2}$, $\ell_c=l^3_c$ and $\ell_{gc}=l^3_{gc}$.
These functions, as well as expressions for the asymptotic behaviour
for $\omega \rightarrow \infty$ can be found in~\cite{critexp}. In terms of
the above, the threshold function appearing in $\beta_{e^{2}}$
reads
\begin{eqnarray}
{\ell}^d_g(2 \lambda \kappa,
2 e^2 \kappa ) \!&=&\! \frac{48 e^2 \kappa}{d (d+2)} \left[
(d+1) m^d_{2,2}(2 \lambda \kappa, 2 e^2 \kappa) -
n^{d-2}_{2,1}(2 \lambda \kappa, 2 e^2 \kappa) \right]\nonumber\\
&&+\frac{m^{d+2}_{2,2}(2 \lambda \kappa, 0)}{m^{d+2}_{2,2}(0,0)}~l^d_g.
\label{A.2}
\end{eqnarray}
%
If one includes the $\rho$-dependence of the
kinetic term one needs modified threshold functions as
\begin{eqnarray}
\tilde\ell^d_0(\omega,\gamma)&=&{1\over2}~k^{-d}
\int_0^\infty dx~x^{\frac{d}{2}-1}
\tilde\partial_t \ln(P(x)+\omega k^2+(\gamma-1)~x) \\
\tilde\ell^d_n(\omega,\gamma)&=&-{1\over2}~k^{2n-d}
\int_0^\infty dx~x^{\frac{d}{2}-1}
\tilde\partial_t (P(x)+\omega k^2+(\gamma-1)~x)^{-n}
\end{eqnarray}
with similar replacements in the other threshold functions.

\end{appendix}
\bigskip

\newpage
\section*{\names Figure Captions}
\begin{description}
\item [Fig.~1:] The Feynman graph corresponding
to the contribution $\sim e^4$
in equation (\ref{lambdaflow}).
\item [Fig.~2:] The phase diagram corresponding to the evolution
eqs.~(\ref{kappaflow}-\ref{e2flow}).
We present the projection for $e^2=e^2_\star$ and the arrows
indicate the flow with $k\to0$ (see text).
\item [Fig.~3:]  Same as Fig.~2, but the projection
 is taken for $\kappa=\kappa_c(e^2,\lambda)$.
\item [Fig.~4:] Same as Fig.~3 near the gaussian fixed point.
\item [Fig.~5:] The $\beta$-function for $\lambda$ on the critical surface.
The inset shows the region for very small $\lambda$.
\item [Fig.~6:] The $\beta$-function for $\lambda/e^2$ in the linear
approximation, eq.~(\ref{betaratio}). One observes no fixed points
for $\lambda/e^2$: $\beta_{\lambda/e^2} \geq 0.13~e^2$.
\item [Fig.~7:] The phase diagram of the effective scalar
 theory in the limit $e^2\rightarrow\infty$. The
 region where
 $e^2\rightarrow\infty$ on the critical line is denoted by crosses,
the one where $e^2$ decreases for $k\rightarrow 0$
by diamonds.
\item [Fig.~8:]  Expected curve of
$\partial \tilde u/\partial r$, eq.~(\ref{g75}),
near the origin. The non-trivial extrema
of the potential corresponds to the
intersections of the curve with the
dashed line. a) $l < (4-2k_{\rm tr}/\Lambda)A$.
For the value of $\sigma$ depicted here it shows two non-trivial
extrema: a maximum, close to the origin, and a
minimum. b) $l > (4-2k_{\rm tr}/\Lambda)A$.
In this case the potential has no
local maximum for any value of $\sigma$. The figure depicts a value
of $\sigma$ showing a minimum.
\item [Fig.~9:]  The dashed line indicates the border between
negative (above) and positive (below)
$\eta_{\scriptscriptstyle F}$ for $N=1$. In addition,
the fixed point values for the tricritical (left branch) and the
second-order (right branch) fixed points are given for various values of $N$,
corresponding to the $\varphi^4$- (stars)
and the $\varphi^8$- (diamonds) approximation.
\end{description}

\eject
\vfill

\epsffile{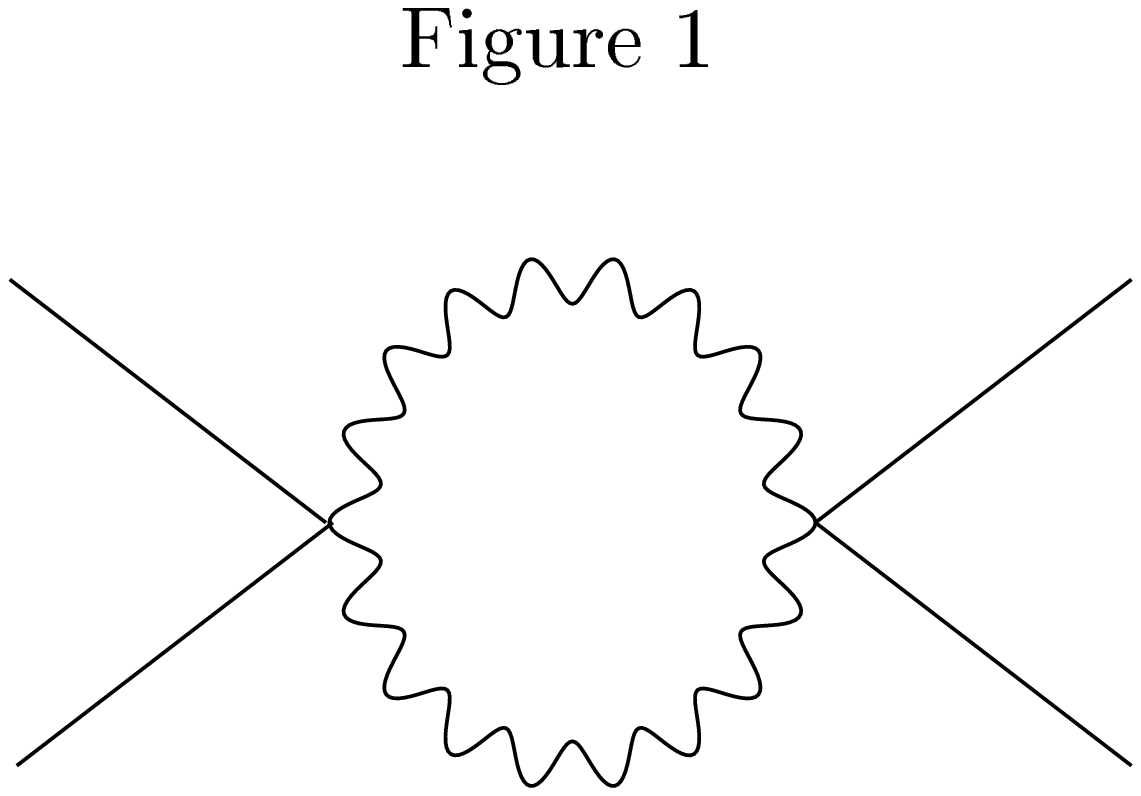}
\epsffile{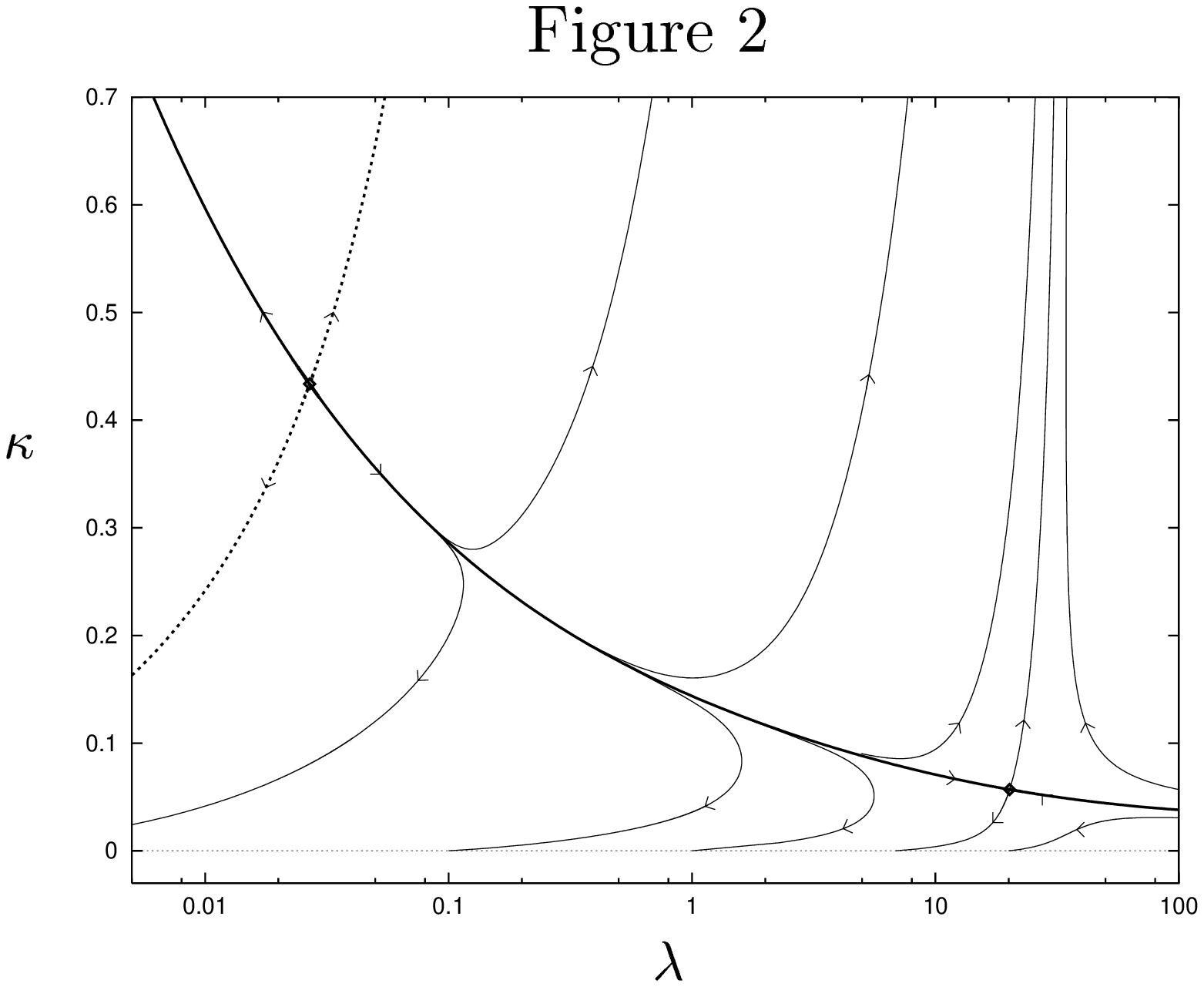}
\epsffile{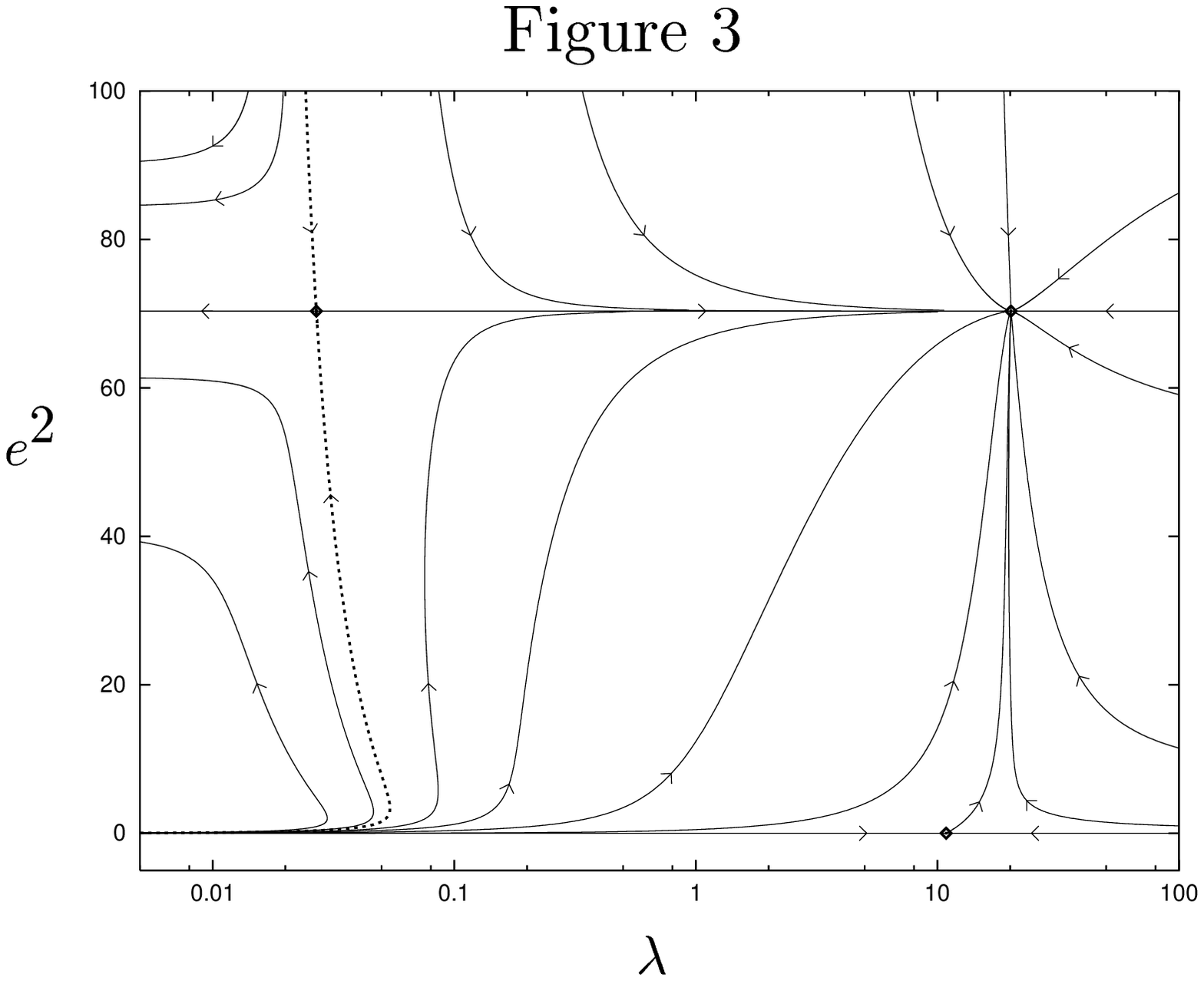}
\epsffile{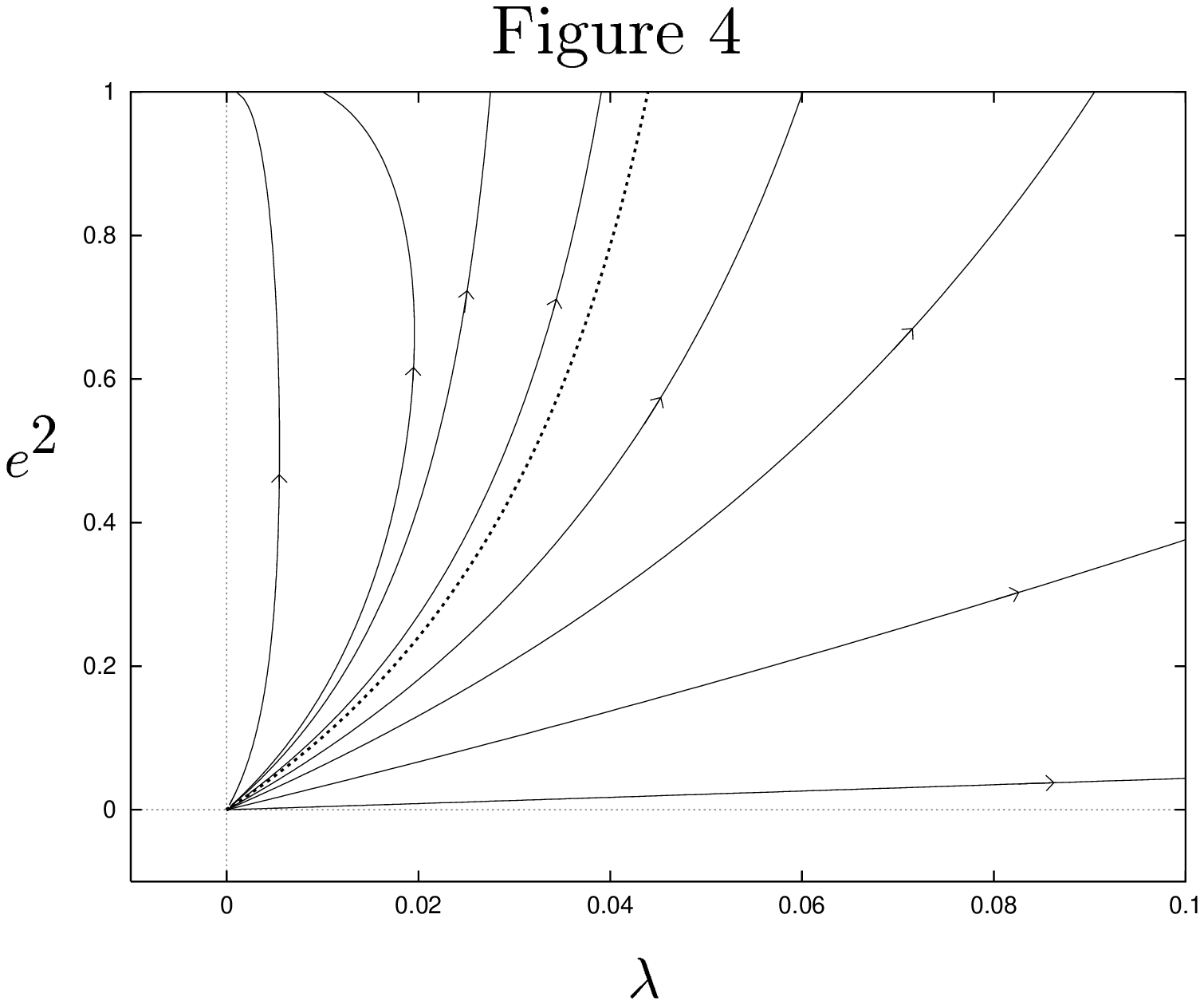}
\epsffile{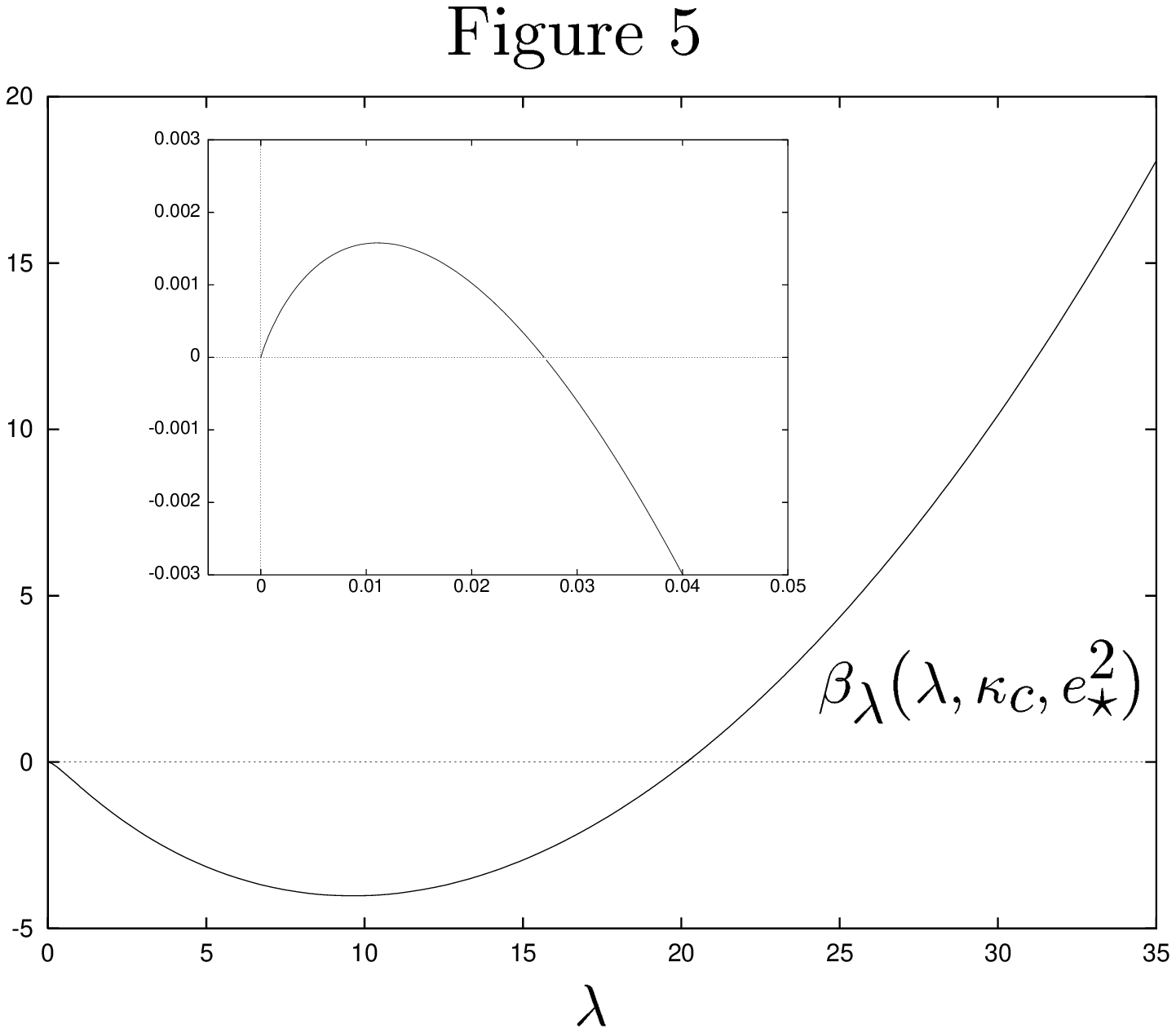}
\epsffile{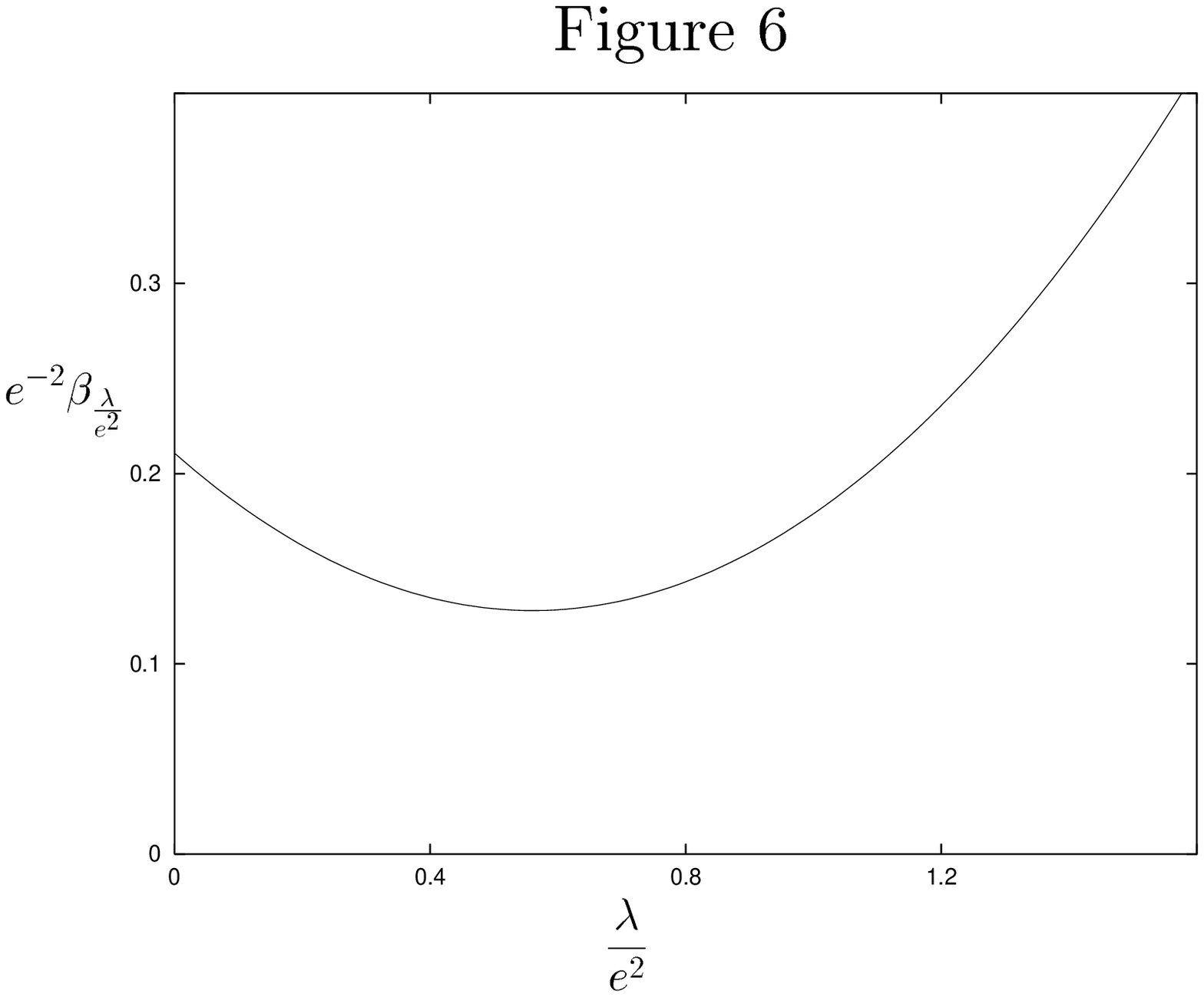}
\epsffile{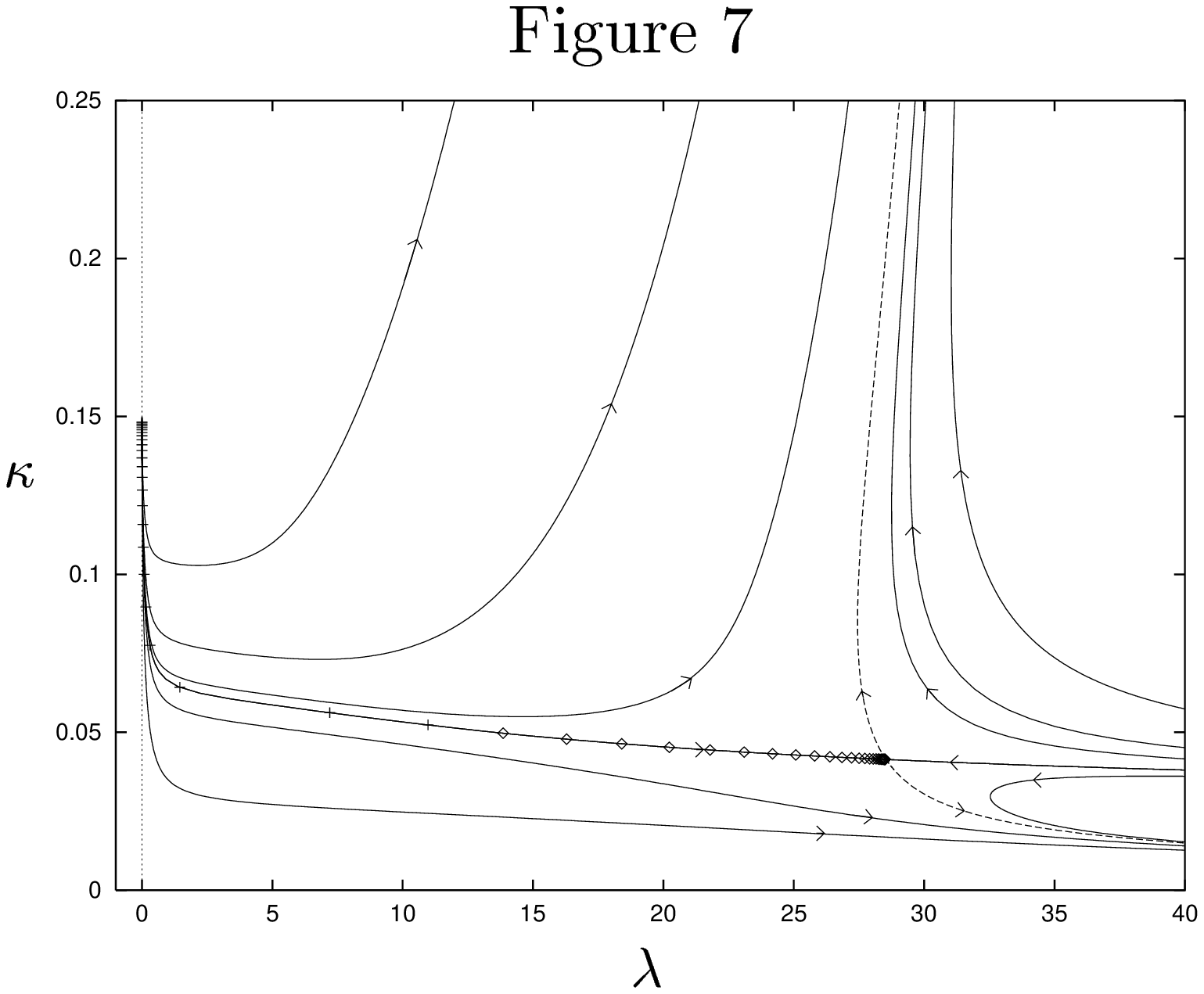}
\epsffile{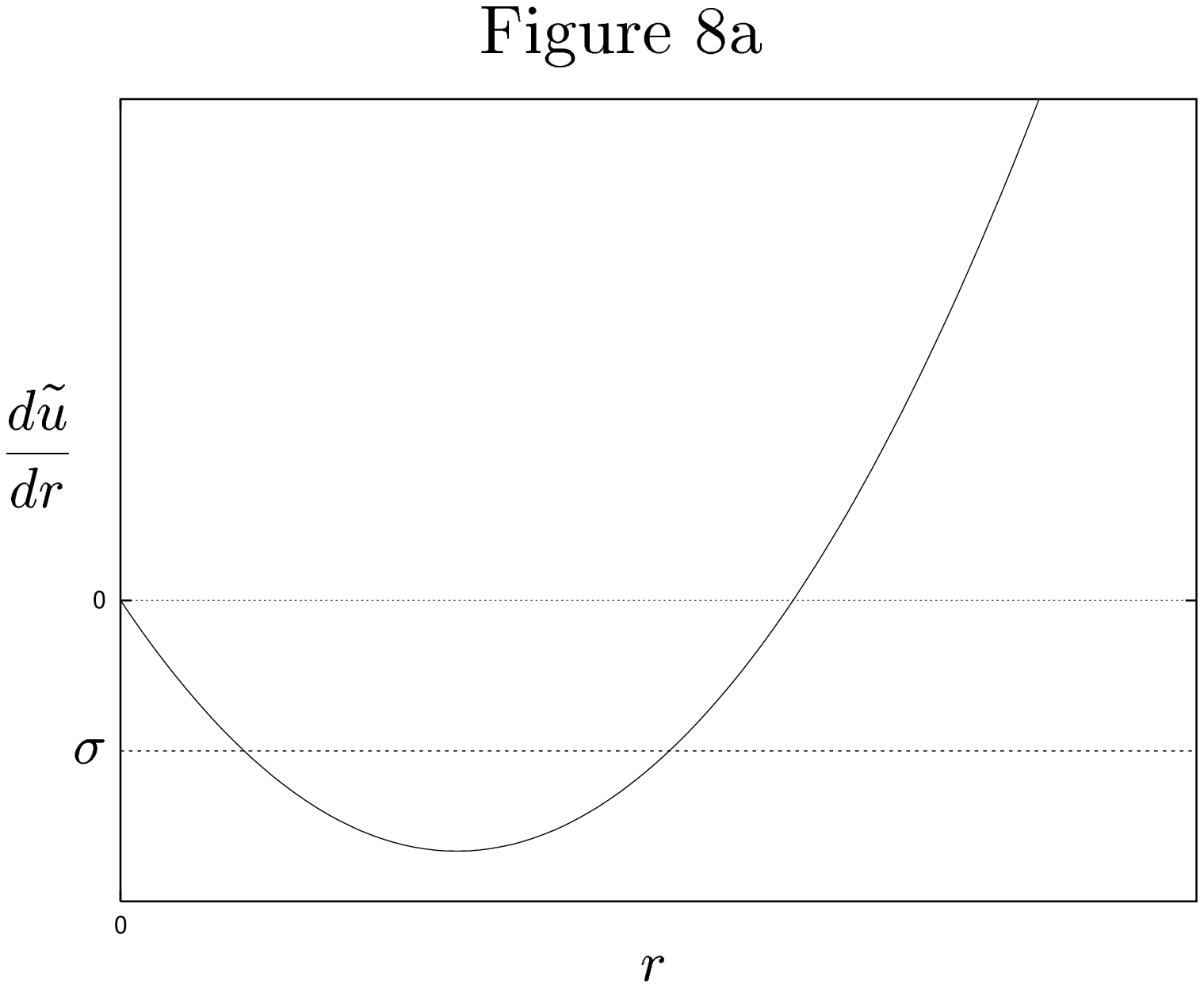}
\epsffile{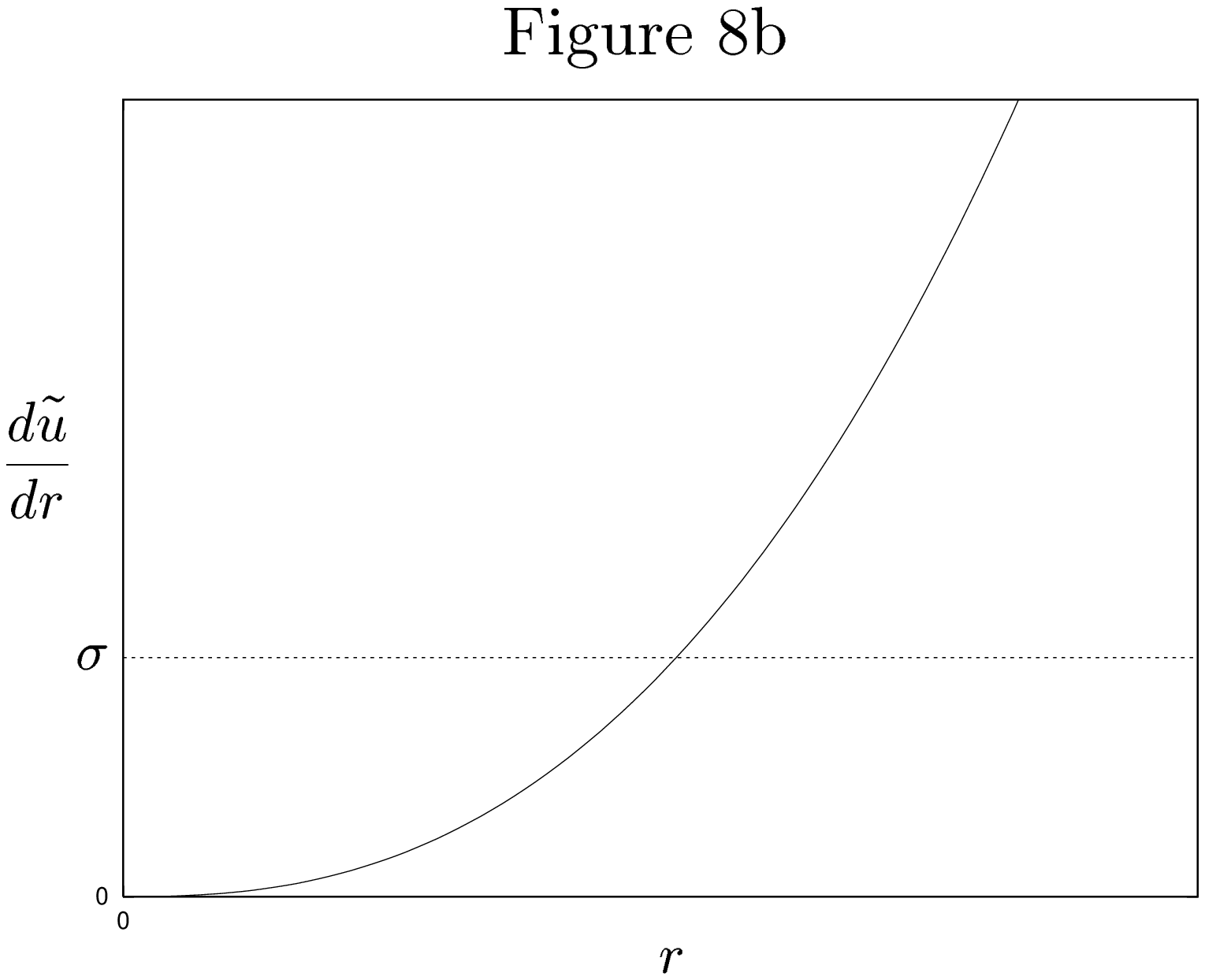}
\epsffile{newfig9.eps}


\begin{thebibliography}{99}

\bibitem{HLM}
B.\,I.\,Halperin,
T.\,C.\,Lubensky and S.\,Ma,
Phys. Rev. Lett. {\bf 32}
(1974) 292.


\bibitem{DH}
C. Dasgupta and B. I. Halperin, Phys. Rev. Lett.
{\bf 47} (1981) 1556;\\
J.\,Bartholomew, Phys.\,\,Rev. {\bf B28} (1983) 5378.


\bibitem{noepsilon}
H.\,Kleinert, Lett.\,Nuovo Cimento {\bf 35} (1982) 405; Gauge
Fields in Condensed Matter (World Scientific, Singapore, 1989);\\
J.\,March-Russell, Phys.\,\,Lett. {\bf B296} (1992) 364;\\
M.\,Kiometzis, H.\,Kleinert and A.\,M.\,J.\,Schakel,
Phys.\,\,Rev.\,\,Lett. {\bf 73} (1994) 1975.


\bibitem{rad}
L.\,Radzihovsky, Europhys.\,\,Lett. {\bf 29} (1995) 227.


\bibitem{hightc}
M.\,B.\,Salomon {\it et al.}, Physica {\bf A200} (1993) 365.


\bibitem{gennes}
P.\,G.\,de Gennes, Solid State Commun. {\bf 10} (1972) 753;\\
B.\,I.\,Halperin and T.\,C.\,Lubensky, Solid State Commun.
{\bf 14} (1974) 997.


\bibitem{nemdata}
D.\,L.\,Johnson {\it et al.}, Phys.\,\,Rev. {\bf B18}
(1978) 4902;\\
C.\,W.\,Garland, G.\,B.\,Kasting and K.\,J.\,Lushington,
Phys.\,\,Rev.\,\,Lett. {\bf 43} (1979) 1420;\\
C.\,W.\,Garland {\it et al.}, Phys.\,\,Rev. {\bf A27}
(1983) 3234;\\
J.\,Thoen, H.\,Marynissen and W.\,van Daal,
Phys.\,\,Rev.\,\,Lett. {\bf 52} (1984) 204; \\
B.\,M.\,Ocko {\it et al.},
Phys.\,\,Rev.\,\,Lett. {\bf 52} (1984) 208.

\bibitem{newdata}
C.\,W.\,Garland and G.\,Nounesis, Phys.\,\,Rev.
{\bf E49} (1994) 2964, and references therein.

\bibitem{CW}
C. Wetterich, Phys. Lett. {\bf B301} (1993) 90.


\bibitem{CW1}
C.\,Wetterich, Nucl.\,\,Phys. {\bf B352}
(1991) 529;\\
C.\,Wetterich, Z.\,\,Phys. {\bf C57}
(1993) 451; {\bf C60} (1993) 461.


\bibitem{NPB427}
M.\,Reuter and C.\,Wetterich,  Nucl.\,\,Phys. {\bf B408} (1993) 91;
{\bf B427} (1994) 291.


\bibitem{polchinski}
F.\,Wegner and A.\,Houghton, Phys.\,\,Rev.\,\, {\bf A8} (1973) 401;\\
K.\,G.\,Wilson and I.\,G.\,Kogut, Phys.\,\,Rep.\,\, {\bf 12} (1974) 75;\\
S.\,Weinberg, Critical Phenomena for Field Theorists,
Erice Subnucl. Phys. (1976) 1;\\
J.\,Polchinski, Nucl.\,\,Phys.\,\, {\bf B231} (1984) 269.


\bibitem{critexp}
N.\,Tetradis and C.\,Wetterich,
Nucl.\,\,Phys. {\bf B422} (1994) 541.


\bibitem{scalar_AEA}
M.\,Gr{\"a}ter and C.\,Wetterich,
Phys.\,\,Rev.\,\,Lett. {\bf 75} (1995) 378;\\
J.\,Berges, N.\,Tetradis and C.\,Wetterich, Heidelberg preprint
HD-THEP-95-27.

\bibitem{largeN}
B.\,Bergerhoff, D. Litim, S. Lola and
C. Wetterich, Phase transition of N-component superconductors,
Heidelberg preprint HD-THEP-94-49.


\bibitem{epsilonexp}
J.-H.\, Chen, T.\,C.\,Lubenski and D.\,R.\,Nelson,
Phys.\,\,Rev. {\bf B17} (1978) 4274;\\
I.\,D.\,Lawrie, Nucl.\,\,Phys. {\bf B200}[FS4] (1982) 1;\\
S.\,Kolnberger and R.\,Folk, Phys.\,\,Rev.\,\, {\bf B41} (1990) 4083;\\
P.\,Arnold and L.\,G.\,Yaffe, Phys.\,\,Rev.\,\, {\bf D49} (1994) 3003.


\bibitem{abbott}
L.\,F.\,Abbott, Nucl.\,\,Phys.\,\, {\bf B185} (1981) 189.


\bibitem{goldenfeld}
see\,{\it {e.g.}~}\,N.\,Goldenfeld, Lectures on Phase Transitions and
the Renormalization Group (Addison-Wesley, Reading, MA, 1992).

\bibitem{Ellwanger}
U.\,Ellwanger, Phys.\,\,Lett. {\bf B335} (1994) 364; \\
U.\,Ellwanger, M.\,Hirsch and A.\,Weber, Orsay preprint
LPTHE-ORSAY-95-39.


\bibitem{DE}
T.\,R.\,Morris, Phys.\,\,Lett. {\bf B329} (1994) 335;
preprint CERN-TH 7403/94.


\bibitem{NPB391}
M.\,Reuter and  C.\,Wetterich,
Nucl.\,\,Phys. {\bf B391} (1993) 147.


\bibitem{cpn}
M.\,L\"uscher, Phys.\,\,Lett. {\bf B78} (1978) 465;\\
E.\,Witten, Nucl.\,\,Phys. {\bf B149} (1979) 285.




\bibitem{nematic}
S.\,Hikami, Prog.\,Theor.\,Phys. {\bf 62},\, No.1 (1979) 226.


\bibitem{private}
N.\,Tetradis, private communication.


\bibitem{4D}
D.\,Litim, N.\,Tetradis and C.\,Wetterich,
Non-Perturbative Analysis of the
Coleman-Weinberg Phase Transition, preprint HD-THEP-94-23 / OUTP-94-12.
\end{thebibliography}
\end{document}